\documentclass{cernrep}
\usepackage{hyperref}
\usepackage{tcolorbox}
\usepackage{graphicx,subfigure}
\usepackage[mode=match]{siunitx}

\usepackage[noblocks]{authblk}

\renewcommand{\thefootnote}{\fnsymbol{footnote}}
\begin{document}

\title{{\vspace*{-18mm}\normalfont\small CERN-LPCC-2025-002}\\[6mm]
{\huge Reinterpretation and preservation of data and analyses in HEP}\vspace*{6mm} \hrule \vspace*{3mm}
\parbox{15.8cm}{\normalfont\normalsize Data from particle physics experiments are unique and  are often the result of a very large investment of resources. Given the potential scientific impact of these data, which goes far beyond the immediate priorities of the experimental collaborations that obtain them, it is imperative that the collaborations and the wider particle physics community publish and preserve sufficient information to ensure  that this impact can be realised, now and into the future. The information to be published and preserved includes the algorithms, statistical information, simulations and the recorded data. This publication and preservation requires significant resources, and should be a strategic priority with commensurate planning and resource allocation from the earliest stages of future facilities and experiments.}\\[4mm] \hrule 
\vspace*{2mm}
{\normalfont\small Document submitted to the European Strategy for Particle Physics Update 2026.}\\[1mm] {\normalsize\emph{LHC Reinterpretation Forum:}} \vspace*{-5.8mm}}

\footnotetext[2]{editor}
\footnotetext[1]{contributor}
\thispagestyle{empty}


\author[1]{Jonathan~Butterworth$^\dagger$}
\affil[1]{University College London, UK}

\author[2]{Sabine~Kraml$^\dagger$}
\affil[2]{LPSC Grenoble, Universit\'e Grenoble-Alpes, CNRS/IN2P3, France}

\author[3]{Harrison~B.~Prosper$^\dagger$}
\affil[3]{Florida State University, USA}


\author[4]{Andy~Buckley$^*$}
\affil[4]{University of Glasgow, UK}

\author[5]{Louie~Corpe$^*$}
\affil[5]{LPCA Clermont, Universit\'e Clermont-Auvergne, CNRS/IN2P3, France}

\author[6]{Cristinel~Diaconu$^*$}
\affil[6]{CPPM, Aix Marseille Univ., 
CNRS/IN2P3, France}

\author[7]{Mark~Goodsell$^*$}
\affil[7]{LPTHE, Sorbonne Universit\'{e} \& CNRS, Paris, France}

\author[8]{Philippe~Gras$^*$}
\affil[8]{IRFU, CEA, Universit\'e Paris-Saclay, 
France}

\author[4]{Martin~Habedank$^*$}
\affil[4]{University of Glasgow, UK}

\author[9]{Clemens~Lange$^*$}
\affil[9]{Paul Scherrer Institute, Switzerland}

% ...

\maketitle



\renewcommand{\thefootnote}{\arabic{footnote}}
\pagenumbering{arabic} 

\section{Introduction}

Experimental results in particle physics are unique and of immense scientific value.
They are obtained at enormous cost in terms of both financial and human resources, and their implications reach far beyond the sets of calculations, theories or parameter combinations with which they are confronted in the original experimental publications.
Indeed, their implications may reach well beyond any theory brought forth so far and thus beyond anything that may directly be tested by the experimental collaborations themselves.
It is thus crucial that their results can be used and interpreted by the whole physics community --- now and for decades to come.

The products of scientific research are multiple and extend far beyond the traditional mode of journal-based results publication.
We call for action to make the preservation and publication of these various outputs normative through suitable long-term public storage.
Analysis-specific data products and related algorithms, software and workflows can only be preserved if the work is done at the time of active analysis.
Supporting these practices and funding domain-specific services to make them available is hence crucial.
These products have a potential for an immediate reuse and feedback to the experimental collaborations, as well for active
exploitation beyond the lifetime of any given experiment.

While there is broad agreement on the importance of information preservation, and detailed recommendations have been worked out over the
years~\cite{Kraml:2012sg,LHCReinterpretationForum:2020xtr,Cranmer:2021urp,Bailey:2022tdz,Bailey:2022pdq,Araz:2023mda}, the situation nevertheless remains dire, with significant loss of information, and therefore of physics, already occurring in current experiments.
Once an analysis has been published and the main analysers (typically Ph.D.\ students or postdocs) have moved on, 
it is often impossible to reproduce the analysis with high fidelity even within the collaboration that produced the work, or to retrieve any missing information
for further studies, including reinterpretation. Restrictive collaboration policies on the release of information, including simulated data samples,
analysis code or detector calibrations,
can make the external use of such information even more challenging.

Current practice severely limits the shelf-life and scientific impact of many published analyses. The problem affects not only further studies
by collaboration outsiders, often theorists, but even the experimental collaborations themselves.
Indeed, the reproducibility of 
several LHC Run-1 and even Run-2 searches for new physics have already been lost within ATLAS and CMS \cite{RIFtalks}!

Therefore, it is crucial that
\emph{experimental collaborations in\-te\-grate extensive data- and analysis-preservation efforts into their analysis and publication processes} from the beginning,
facilitating communication of analyses and results in reusable form. 
\emph{These initiatives need to be supported by the host laboratory, not only by acknowledging Open Science as one of its core missions but also by allocating sufficient resources for Open Science services and Open Data storage.}

In the rest of this contribution, we expand on what kinds of information are required and for what purposes and provide some evidence to support
this \textit{cri de cœur} in the form of examples of the scientific impact of the laudable but limited efforts so far made.
We draw on previous efforts to make this case, and in particular on the relevant Snowmass white papers \cite{Bailey:2022tdz,Bailey:2022pdq}.

\section{What public, preserved information is needed?}


\noindent
To fully exploit and future-proof our experiments, the results and other materials that need to be preserved and made available
to the whole community include:
\begin{itemize}
  \item \textbf{Event-level data}: data recorded from experiments; outputs from passing those data through
    the experiment's standard event reconstruction, and analogous simulated data samples. This may
    include data that have been passed through a data compression step that prunes
    information, but which might also add additional quantities to the data.
  \item \textbf{Generator-level data}: simulated events that have not been passed through detector simulations.
  \item \textbf{Analysis data products}: selections of derived data and synthesised information from the various stages of an analysis. These might include histograms of kinematic distributions, fiducial cross-sections, cross-section limits,
    simplified model results, theoretical predictions used for interpretation of the data, correlation information, likelihoods, etc.
  \item \textbf{Analysis code and workflows}: the software infrastructure that implemented the original analysis. This  encompasses the complete analysis workflow, including accessing, reformatting, and reducing data; executing the analysis physics logic; accessing and incorporating various external metadata (including theoretical input such as cross sections, weights, or experimental input such as resolutions, scale factors, etc.), and obtaining the analysis data products, packaged to enable the exact reproduction of the analysis. 
  \item \textbf{Analysis logic}: 
    the essential information required to understand and reproduce an analysis and associated analysis data products. This needs to be provided in sufficient detail for reproduction and future work, ideally in programmatic form, ensuring clarity, accessibility, and reproducibility. 
  \item \textbf{Detector-performance data}: either the data need to    be corrected for detector performance to within 
    quantified uncertainties
    (`unfolded'), or sufficient information needs to be provided to emulate the impact of detector effects on the true observables.
  \item \textbf{Statistical models}: the full model describing the probabilistic dependence
    of the observable (and observed) data on the parameters of interest and the nuisance parameters. This is essential information for the accurate statistical evaluation of any new signal hypothesis in terms of parametric or kinematic reinterpretations. Moreover, it enables updates of an analysis when more precise theoretical predictions or improved experimental calibrations become available.
\end{itemize}

Although many physics applications can succeed with subsets of the above information, we argue that the aspiration
should be to be as complete as possible in the information that is made publicly available and preserved, to maximize the openness and reproducibility of our science.
Publicly released material should follow the data management principles of
\textit{Findable, Accessible, Interoperable, and Reusable} (FAIR) as described in Ref.~\cite{FAIR:2016}. This will require concerted attention and resources throughout the end-to-end life cycle of data generation,
processing, analysis, preservation, and distribution, and should be considered already in the  design and commissioning phase of experiments.     
This is closely connected to the need for long-term software transparency and sustainability.\footnote{Many large experiments have made their software codebase publicly available and version controlled, which is a welcome development towards being transparent. However, preserving the code alone does not satisfy the requirements mentioned so far and does not cover the whole analysis chain.}

\section{Current practice and impact}

\subsection{Event-level data}

\begin{figure}[ht]
  \centering \includegraphics[width=0.6\textwidth]{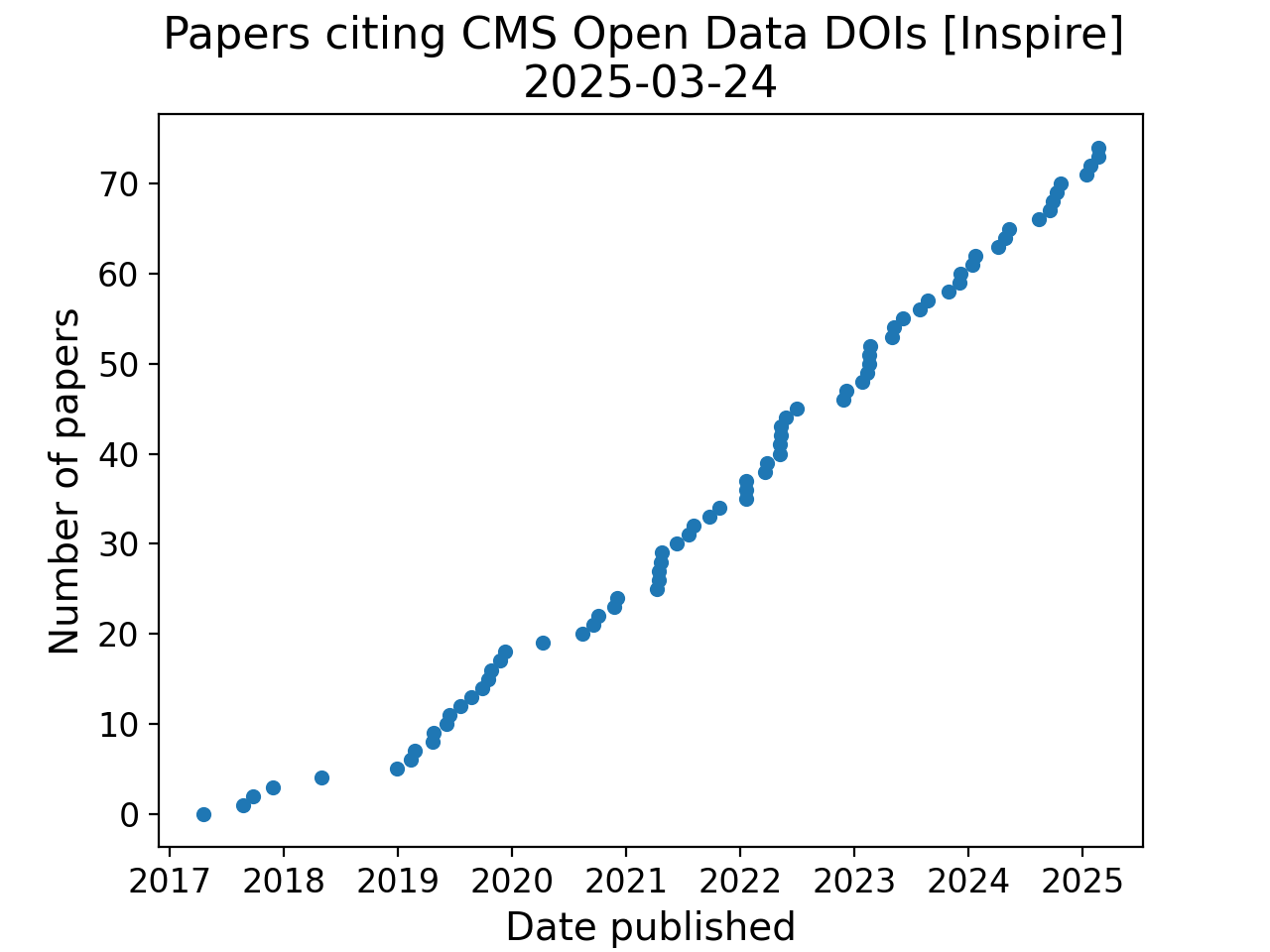}
  \caption{ 
  The number of publications citing CMS Open Data records, 
  excluding CMS collaboration and CMS Open Data contributors (source: \cite{mccauley_2025_15078671}). 
    \label{fig:CMSOpenData}}
\end{figure}

All LHC experiments are committed to releasing research-quality data --- collision data and simulations ---
through the CERN Open Data portal~\cite{CERN-OPEN-DATA}.
The amount of data and release timeline varies from one experiment to another~\cite{CERN-OPEN-2020-013}.
Experience from the pioneering CMS Open Data demonstrates that once data are available and their usage well documented, new original research is inspired as shown in Fig.~\ref{fig:CMSOpenData}.

CMS Open Data have been used in physics studies, benchmarking research software infrastructure, data-science applications, and as a proof of principle for scientific methods.  
Authors appreciate the authenticity of these data: e.g.,\ the presence of experimental uncertainties
and noise, as well as the availability of simulations, mandatory for research-level physics studies.

The event-level data thus far released by the experimental collaborations come with some limitations. For example, they typically do not include enough information to develop low-level reconstruction algorithms or to perform complex long-lived particle analyses. Bespoke, limited-statistics datasets can satisfy the former use case, and the collaborations should interact with and monitor the interests of the broader HEP community to identify necessary targeted datasets. For the latter use case, more work is needed to maximise the utility of the publicly released data formats to ensure maximal physics breadth in data analysis based on open data.

When the LHC enters its High Luminosity phase, data volumes and resource challenges for open data initiatives will increase significantly.
A common understanding that these open data represent the scientific heritage of the LHC and the only available high-energy hadron collision data for decades to come is essential to secure the necessary resources beyond Run-3 data.
The amount of open data is approximately 5~PB at the moment, and it is expected to grow with increasing integrated luminosity, with a commensurate growth in simulated data.  While the amount of open data at the petabyte scale is large, the associated cost of preserving it is minimal compared with the overall cost of LHC operations.\footnote{Note here that the open data effort comes on top of (internal) data preservation, which requires experts in charge of data integrity and stability, who are able to rerun the experimental original software and reproduce/reconstruct data as discussed in~\cite{DPHEP:2023blx} and the corresponding contribution by the DPHEP collaboration to the ESPPU~\cite{DPHEP:ESPPU26}.}

\subsection{Generator-level data}
 
All LHC collaborations, as well as collider phenomenologists,  simulate large Standard Model (SM) event samples using the same set of publicly available tools. Significant duplication of effort and resource consumption can be mitigated by the sharing of simulated events across the HEP community.  Centralising the production of SM event samples to a core LHC team working in consultation with Monte Carlo experts should be prioritised as a strategic objective, resulting in substantial financial and environmental savings and collateral benefits in increased robustness, transparency, and equitable access.
This will require agreement to align the event generation workflows between different experiments.  

Effective curation of these data, and their release in a format that is compatible with a large set of downstream phenomenological
tools is essential to ensure their uptake within the community.  We note that the ATLAS collaboration has swiftly responded to this need by committing to a first release of \textsc{HepMC} format~\cite{Buckley:2019xhk}) events within a timeframe of order 1-2 months~\cite{Marshall2025}. 

\subsection{Analysis data products}

\begin{figure}[!t] \centering
    \includegraphics[width=0.65\textwidth]{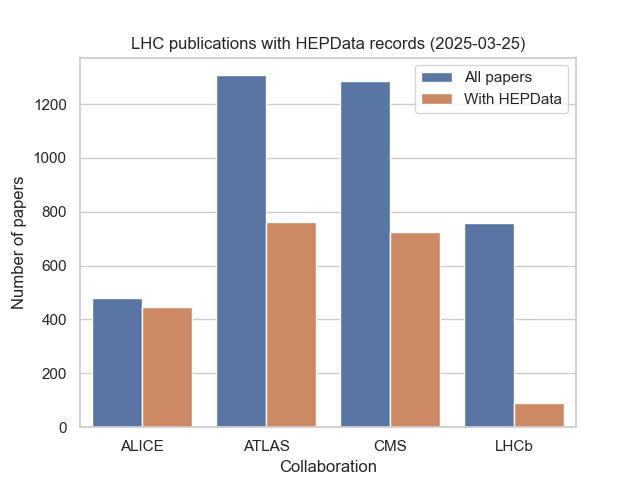}
    \caption{Number of publications by the four main LHC collaborations compared to publications with associated HEPData records (source: \cite{HEPData:notebook}).\label{fig:HEPData}}
\end{figure}

The standard repository for digitised, reusable data products is HEPData~\cite{Maguire:2017ypu,Buckley:2010jn}, which
preserves more than \mbox{10,000} publication records 
containing (differential) cross-section measurements and limits,
correlation information, associated theoretical predictions and other derived data from hundreds of past and current experiments. However, as shown in Fig.~\ref{fig:HEPData}, only a fraction of the hundreds of publications
from LHC experiments provide reusable material on HEPData. 
This is extremely unfortunate, as it severely limits the scientific impact of the associated publications. 

We also note that publication records submitted to HEPData need to be complete enough  to be useful! 
In particular, all plots that document an analysis and its performance need to be preserved in digital form, 
not only a subset thereof. Likewise, providing details is important; for example, it is extremely useful if all components of a histogram (including, e.g., the detailed breakdown of SM expectations) are made available. 
Finally, additional auxiliary material is often needed to enable correct reuse, beyond the plots and tables presented in the paper.\footnote{The (re)use cases are manifold. For instance, besides the simulation-based reinterpretation tools mentioned in section~\ref{sec:logic}, tools and efforts like SModelS~\cite{Kraml:2013mwa,Ambrogi:2017neo,Ambrogi:2018ujg,Alguero:2021dig,MahdiAltakach:2023bdn,Altakach:2024jwk}, HiggsTools~\cite{Bechtle:2008jh,Bechtle:2011sb,Bechtle:2013xfa,Bechtle:2013wla,Bechtle:2020pkv,Bechtle:2020uwn,Bahl:2022igd}, Darkcast~\cite{Ilten:2018crw}, SMEFit~\cite{Giani:2023gfq} and many others critically rely on the provided material.}   
Detailed recommendations are given in Ref.~\cite{LHCReinterpretationForum:2020xtr,Bailey:2022tdz}. 
As noted earlier,  missing information is very difficult to retrieve, as often the analysis team has disbanded with analysers leaving the field, or the relevant files become lost. 
This problem could likely be ameliorated if the provision of (auxiliary) material on HEPData were part of the standard publication pipeline, with appropriate infrastructure support. 
The \href{https://indico.cern.ch/category/14155/}{RAMP} (``Reinterpretation: Auxiliary Material Presentation'') seminar series aims at providing more visibility and recognition for such efforts,  but more work is needed 
to make the preservation of all necessary data products on HEPData standard practice.

Another aspect of the reuse problem is that data submitted to HEPData are primarily in tabular form. 
Material beyond digitised plots (like Monte Carlo run cards, input files for benchmark points, but also statistical or machine-learned models) are ``additional resources'', often lumped together in compressed archives without any standard structure.  To accommodate 
the increasingly rich and diverse data products produced in HEP, the HEPData infrastructure needs to be extended, so that all items are searchable, findable and interoperable in an automated manner.  
 
\subsection{Analysis code and workflows}  

Reproducing an analysis in its entirety requires more than just a description of its methodology, which might
be given in a journal paper. It requires access to the software infrastructure that was used to perform the analysis. Analysis code and workflows encode every step of the process, from raw data access to the final extraction of physics results. 
Without access to the complete computational workflow---the analysis pipeline, precise understanding and reproduction of some results, whether within the original collaboration or by external researchers, is impossible.

Ensuring long-term usability of such analysis software requires structured preservation efforts. This includes documenting dependencies, packaging the code in a well-structured repository, and using version control to track changes. However, software evolves, and maintaining ``executability" over time is a significant challenge. To address this, containerization technologies encapsulate the full analysis environment, including software versions, libraries, and dependencies. This approach allows analyses to be re-executed years after publication with identical computational conditions. 
Beyond containerization, workflow management systems such as REANA~\cite{Simko:2018zzz} provide a structured framework for organizing and executing analyses. By defining the entire workflow in a reproducible manner, these systems facilitate both internal validation and external reinterpretation studies. Integrating such tools into standard analysis pipelines will ensure that experimental results remain accessible, reusable, and verifiable long after their initial publication. 

Internal collaboration initiatives such as those in Refs.~\cite{ATLAS:2015wrn, ATLAS:2016dei, ATLAS:2016hks, ATL-PHYS-PUB-2019-032, ATL-PHYS-PUB-2021-020, ATL-PHYS-PUB-2022-045, ATLAS:2024qmx} build on analyses that have successfully been preserved. However, key ingredients to many physics searches are missing, even to the collaborations, so not all searches can be included. We urge the collaborations to continue their ongoing efforts to improve the situation.

\subsection{Analysis logic}
\label{sec:logic}

The analysis logic for LHC event-level data remains complex. The data releases are typically accompanied by rich additional material, including tutorials and guides. Nevertheless it is to be expected that publication-level studies using open data require at least as much time as do analyses within the collaborations. 
From user feedback it is clear that actionable analysis examples -- code, workflow, and environment -- are by far the preferred learning material beyond the quick-start documentation. But extracting such examples from the internal work of collaborations has been challenging. Efforts within collaborations towards full analysis preservation that exploit containerisation technology will greatly benefit the usability of open data and the long-term preservation of analysis knowledge.

Moving away from event-level data analysis, some data products can be used without access to the exact and complete workflow described above. However, for the full exploitation of analysis data products it is crucial that the logic of analyses also be preserved accurately, and ideally in an executable form. This enables, for example, feedback from the theory community about the impact of experimental analyses beyond the models considered in the original collaboration paper, and the testing of new theoretical ideas against the data from several analyses and experiments in a global approach.
It also permits new physics models (either beyond or within the Standard Model) to be tested years after the original analysis was performed.   
It is an essential feature of experimental science that the results from an experiment remain valid and meaningful for testing future theories, and particle physics is no exception despite the intrinsic complexity of the observables and models.
Furthermore, preserving analysis logic is an important way to preserve a key part of the intellectual heritage of the field, which can serve to educate future generations of physicists about crucial scientific work done by predecessors.

In the spirit of open science, 
several public analysis software frameworks have arisen in recent years. 
The Rivet toolkit~\cite{Bierlich:2019rhm} is the clear choice for (differential) measurements, where detector effects have been unfolded to a fiducial phase-space by the experimental collaboration. The {LHC} collaborations have established Rivet-based measurement-preservation programs, and similar efforts are gaining traction at {RHIC} and are being planned at the EIC. 
In parallel, driven by the need of recasting Beyond-the-SM (BSM) searches, the community has been developing various simulation-based reinterpretation frameworks for reconstruction-level analyses. This includes CheckMATE~\cite{Drees:2013wra,Dercks:2016npn}, MadAnalysis5~\cite{Dumont:2014tja,Conte:2018vmg}, HackAnalysis~
\cite{Goodsell:2024aig} and GAMBIT's ColliderBit~\cite{GAMBIT:2017qxg}, as well as detector smearing in Rivet~\cite{Buckley:2019stt}.\footnote{A detailed overview of approaches and public frameworks is given in Ref.~\cite{LHCReinterpretationForum:2020xtr}.}  
Furthermore, the ATLAS SimpleAnalysis~\cite{ATLAS:2022yru} framework communicates analysis logic in the form of code snippets; currently covered are 56 search analyses, primarily searches for supersymmetry~\cite{simpleanalysis:github}. 
We also note ongoing efforts to develop domain specific languages for analysis logic preservation, concretely ADL~(Analysis Description Language)~\cite{Prosper:2022lnf}, which is designed to encode analysis logic in a transparent and framework-independent way, and which has recently seen some traction within CMS.

Unfortunately, the coverage of analyses remains sparse. 
As noted above, not all LHC analyses provide HEPData records, and even fewer provide accompanying analysis code in the form of a 
public recasting module. A search on HEPData for ``\texttt{analysis:(rivet OR madanalysis OR checkmate)}''
gives \mbox{1,117} results as of March 27, 2025, of which more than \mbox{1,000} are Rivet-only.\footnote{SimpleAnalysis implementations cannot currently be searched for on HEPData, but do not change the overall picture.} 
Of the \mbox{1,117} records, 479~are from the four main LHC experiments (245 ATLAS, 142 CMS, 48 LHCb and 44 ALICE). This is just
over 22\% of the HEPData records from the LHC experiments, which as mentioned above is only a fraction of the published results. 
The numbers are lower again for non-LHC experiments, such as heavy-ion or neutrino physics facilities.

\begin{figure}[t] 
    \hspace{-3mm}
    \includegraphics[width=0.54\textwidth]{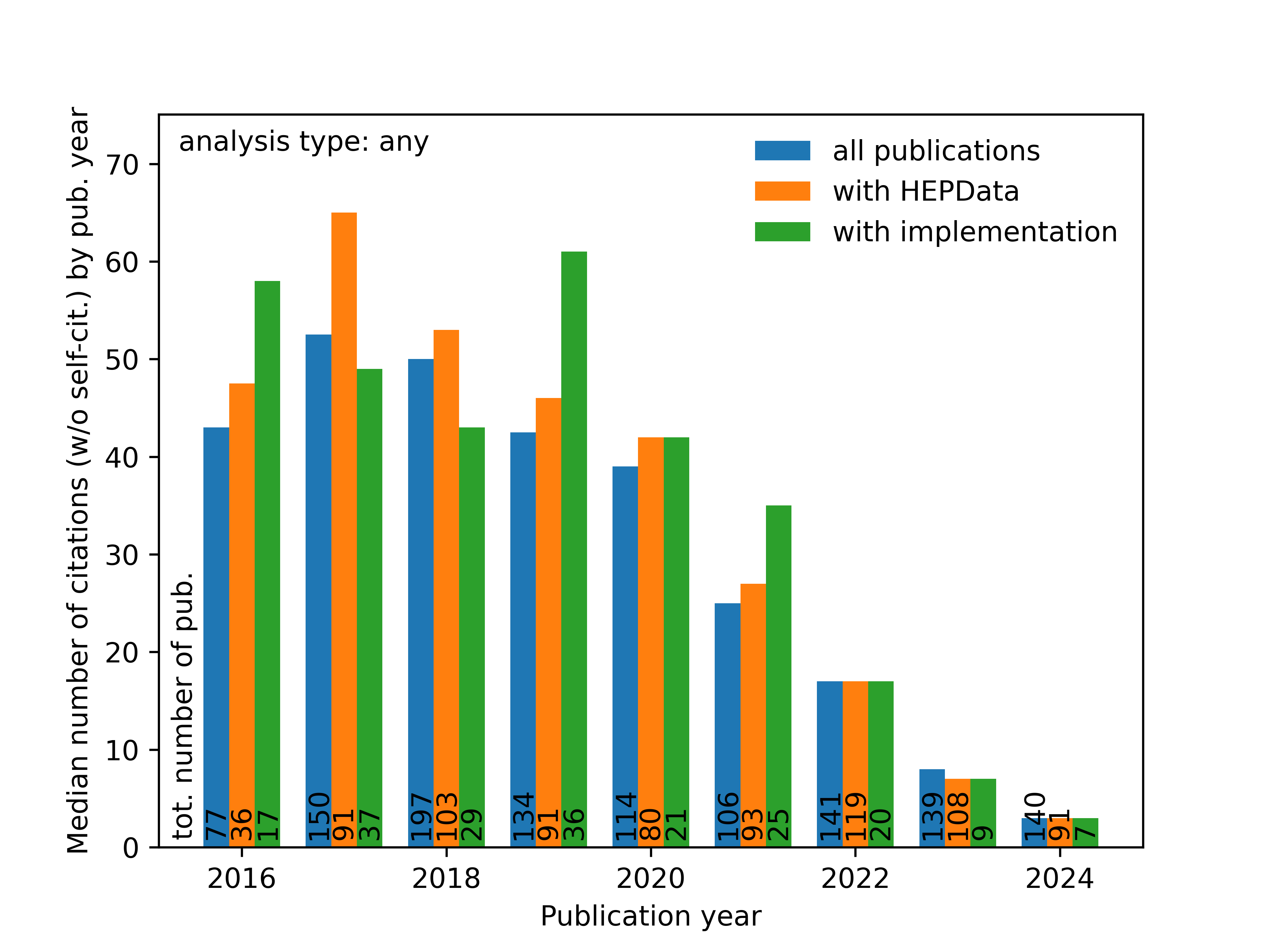}\hspace{-6mm}\includegraphics[width=0.54\textwidth]{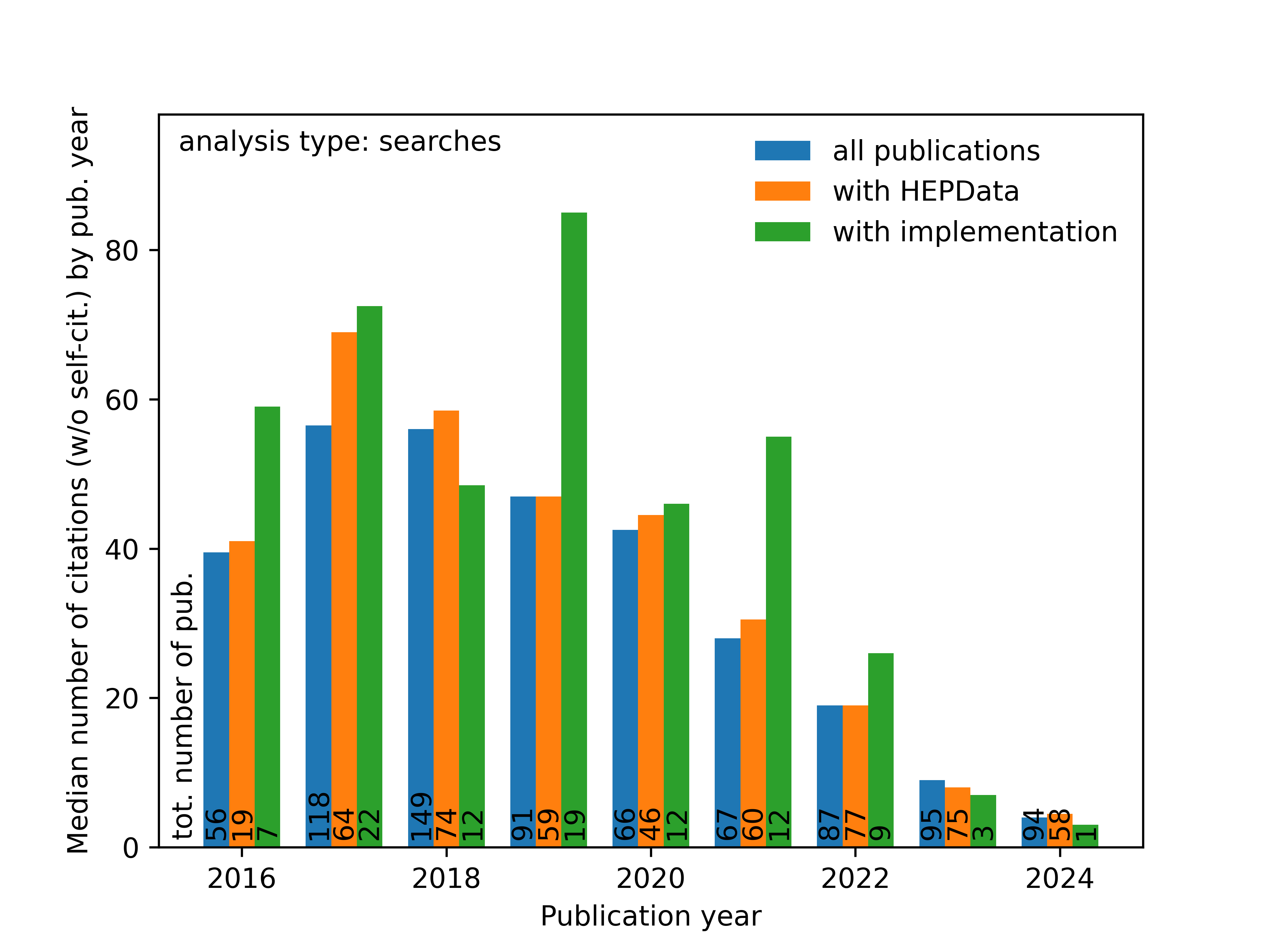}
\caption{%
    Median number of citations (without self-citations) per year for all publications (blue), for publications with a HEPData entry (orange), and, in green, for publications where the analysis logic has been reimplemented in a dedicated tool (CheckMATE, MadAnalysis5, Rivet, or SModelS).
    Given are the numbers for all kinds of ATLAS and CMS analyses at $\sqrt{s}\geq\SI{13}{TeV}$ (left) as well as only for the subset of searches for BSM physics (right). As ``Publication year'' we take the year the paper appeared on the arXiv.  
    (Status 2025-03-13, source:  \cite{habedank_2025_15076716}.) \label{fig:ReinterpretationToolsImpact}}
\end{figure}

One way of quantifying the impact of good practice is to ask how much new research was enabled by making data products and analysis logic available in reusable form to the wider HEP community.
While this is difficult to answer in full generality, the median number of times an experimental publication is cited gives a first indication.
As shown in Fig.~\ref{fig:ReinterpretationToolsImpact}, publications providing a HEPData entry, in particular if it provides sufficient information for a (re)implementation of the analysis logic, get cited more, indicating a higher physics impact.
Overall, publications for which the analysis logic is made accessible are cited \SI{22}{\%} (\SI{62}{\%}) more in the period 2016--2022 (2016--2024).
These values increase to \SI{40}{\%} (\SI{90}{\%}) if searches for BSM physics only are considered, emphasising the fact that preserving the analysis logic is not only imperative for SM measurements.

\begin{figure}[t] \centering
    {\includegraphics[width=0.65\textwidth]{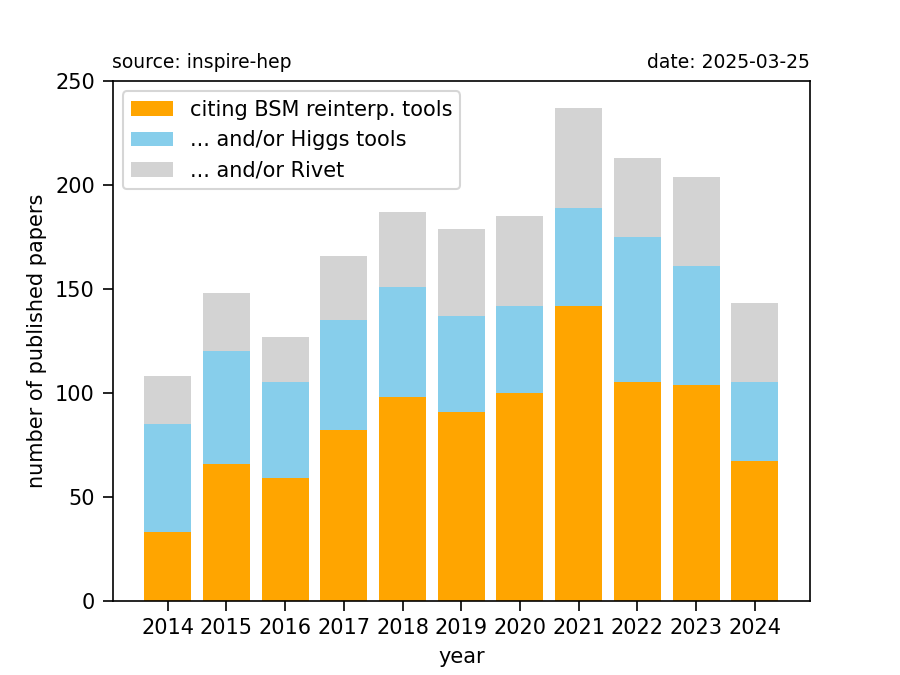}}
\caption{Number of published papers over the last decade making use of or otherwise building upon public BSM reinterpretation tools \cite{Drees:2013wra,Kraml:2013mwa,Conte:2014zja,Dumont:2014tja,Dercks:2016npn,Ambrogi:2017neo,GAMBIT:2017qxg,GAMBIT:2017yxo,Ilten:2018crw,Conte:2018vmg,Ambrogi:2018ujg,Kvellestad:2019vxm,Alvarez:2020yim,Buckley:2021neu,Desai:2021jsa,Lozano:2021zbu,Alguero:2021dig,Araz:2021akd,MahdiAltakach:2023bdn,Goodsell:2024aig,Altakach:2024jwk}, Higgs-specific reinterpretation tools~\cite{Bechtle:2008jh,Bechtle:2011sb,Bechtle:2013xfa,Bechtle:2013wla,Bernon:2015hsa,Kraml:2019sis,Bechtle:2020pkv,Bechtle:2020uwn,Bahl:2022igd}, and/or Rivet~\cite{Buckley:2010ar,Buckley:2019stt,Bierlich:2019rhm,Bierlich:2020wms,Bierlich:2024vqo} (source: \cite{kraml_2025_15084457})\label{fig:ReinterpretationToolsUsage}}
\end{figure}

Another way to illustrate the impact of making more, and better, information public is to trace the impact of the public reinterpretation frameworks that make use of the material provided.
To that end, Fig.~\ref{fig:ReinterpretationToolsUsage} shows the number of published papers since 2014 making use of or otherwise building upon public BSM reinterpretation tools \cite{Drees:2013wra,Kraml:2013mwa,Conte:2014zja,Dumont:2014tja,Dercks:2016npn,Ambrogi:2017neo,GAMBIT:2017qxg,GAMBIT:2017yxo,Ilten:2018crw,Conte:2018vmg,Ambrogi:2018ujg,Buckley:2019stt,Kvellestad:2019vxm,Alvarez:2020yim,Buckley:2021neu,Desai:2021jsa,Lozano:2021zbu,Alguero:2021dig,Araz:2021akd,MahdiAltakach:2023bdn,Goodsell:2024aig,Altakach:2024jwk}, Higgs-specific reinterpretation tools~\cite{Bechtle:2008jh,Bechtle:2011sb,Bechtle:2013xfa,Bechtle:2013wla,Bernon:2015hsa,Kraml:2019sis,Bechtle:2020pkv,Bechtle:2020uwn,Bahl:2022igd}, and/or Rivet~\cite{Buckley:2010ar,Buckley:2019stt,Bierlich:2019rhm,Bierlich:2020wms,Bierlich:2024vqo}. 
It should be stressed that the enabled physics studies encompass a great diversity of new research, from phenomenological studies of models of new physics to new tool developments, from precision calculations to Monte Carlo tuning, and from improved fits of parton densities to novel experimental ideas. While an extensive review is beyond the scope of this contribution, Refs.~\cite{LaCagnina:2023yvi,Baum:2023inl,Araz:2023axv,Chakraborty:2023hrk,Chang:2023cki,GAMBIT:2023yih,Andersen:2023cku,Dreiner:2023bvs,Carrasco:2023loy,Araz:2023bwx,Baer:2023uwo,Arina:2023msd,Bahl:2023xkw,Brahma:2023psr,FASER:2023tle,KA:2023dyz,Coloma:2023oxx,Carpenter:2023agq,Fieg:2023kld,Butterworth:2023rnw,Acaroglu:2023cza,Buonocore:2023kna,Dey:2023exa,Agin:2023yoq,Baer:2023ech,Banerjee:2023upj,Lessa:2023tqc,Frank:2023epx,Acaroglu:2023phy,Carpenter:2023aec,Bagnaschi:2023cxg,Amram:2023jlc,Altakach:2023tsd,DeRomeri:2024dbv,Barman:2024xlc,Gartner:2024muk,Kalinowski:2024uxe,Dehghani:2024lgz,Feike:2024zfz,Chakraborti:2024pdn,Cacciapaglia:2024wdn,Baruah:2024gwy,Argyropoulos:2024yxo,Agin:2024yfs,Heisig:2024xbh,Ellwanger:2024vvs,Butterworth:2024eyr,Balan:2024cmq,Cruz-Martinez:2024cbz,Boto:2024tzp,Corpe:2024ntq,Arbelaez:2024lcr,Butterworth:2024hvb,Fuks:2024qdt,Fan:2024wvo,Athron:2024rir,Lu:2024ade,Cornell:2024dki} may serve as illustrative examples.\footnote{Note that this selection is a subset from 2023--2024 only.}
Finally, we point out that enabling reinterpretation helps i)~identification of interesting benchmark models and signatures not yet covered by current searches, as in \cite{Kim:2014eva,Heisig:2024xbh} and ii)~identifying possible BSM explanations for observed excesses, as in ~\cite{Giudice:2022bpq,GAMBIT:2023yih,Agin:2024yfs}, 
which may impact the design and development of new search strategies.

Before concluding this subsection, a comment is in order regarding analysis logic based on machine learning (ML). 
ML models have a long history in particle physics, but the rapid rise in both relevant computing power and algorithms/architectures in association with wider-world ``AI'' applications has also produced a sea-change in their power and ubiquity in particle physics experiments. While experimentalists are understandably keen to maximally exploit the power of their datasets, trends towards both more complex algorithms and lower-level input features create challenges for preservation and reinterpretation, respectively.
It is therefore important to keep reuse in mind in the analysis design and ensure that the learned model can be deployed in a robust, framework independent form (e.g.\ in ONNX format)~\cite{Araz:2023mda}. 
Related to this it is important for ML-based analysis strategies to consider the appropriate balance between striving for absolutely maximum sensitivity to a specific model, and the use of more universal variables (e.g.\ features at the level of jets or tracks rather than clusters or hits) that enable long-term reuse and reinterpretation.

\subsection{Detector performance data}

Reproducibility of HEP analyses which are not unfolded to particle level critically depends on realistic detector-response modeling. The major LHC experiments already make their simulation and reconstruction software publicly available~\cite{alice-o2,atlas-athena,cms-sw,lhcb-gaudi}. However, running these frameworks is computationally intensive and requires significant expertise, making their direct use challenging for external researchers. Public fast simulation tools, such as DELPHES\,3~\cite{deFavereau:2013fsa}, cited over 3,000 times in the past 12 years, play a key role in enabling phenomenologists and external researchers to explore new physics scenarios and reinterpret experimental results. However, the accuracy of such tools is inherently limited by the availability of reliable detector performance data. To improve public analysis reproduction, experiments should provide well-defined detector response information in a structured and systematic manner.

One approach is to release parameterized or tabulated object-level efficiencies and resolutions, including trigger, reconstruction, and identification efficiencies for key physics objects (e.g., electrons, muons, jets, b-tagged jets, tau leptons, boosted objects, etc.). Publishing these in an accessible format would allow fast simulation tools, or sometimes reinterpretation frameworks themselves, to incorporate them directly and more accurately reproduce detector effects. Another possibility is to provide selected data or Monte Carlo samples that allow the community to develop automated tuning procedures for public fast simulation frameworks, or to train surrogate machine learning models. This would enable data-driven adjustments to improve the fidelity of public packages.

Surrogate models --- ML-based approximations of detector effects, trained to reproduce the output of full simulation and/or reconstruction algorithms using a simpler set of input attributes --- offer a promising and computationally efficient alternative to full simulation while maintaining good accuracy. The provision of such models directly by the experiments as standalone, ready-to-use components would greatly facilitate reinterpretation efforts by external users. A pioneering step in this direction was done by the ATLAS search for displaced hadronic jets~\cite{ATLAS:2024ocv}, which provided the reconstruction-level efficiencies for the full analysis in the form of a BDT surrogate model~\cite{corpe_2024_12957031}, see also~\cite{haddad:ramp}. 
In a similar spirit, the Parnassus (particle-flow neural-assisted simulations) project~\cite{Dreyer:2024bhs}, aims at the automatic construction of surrogate models for combined detector simulation and reconstruction.   

In parallel to the above mentioned options to enable fast and simplified simulation and reconstruction, experiments should continue improving access to their standard simulation and reconstruction software, for example, by using containers and cloud computing, to support more detailed studies when needed.

\subsection{Statistical models}

The statistical model is an appropriate starting point for any detailed interpretation of experimental results~\cite{Cranmer:2021urp} as it
provides a complete probabilistic description of the observable data associated with an experimental analysis.  
When the observed data are entered into the statistical model, the latter becomes the likelihood function of the analysis, which is the key ingredient in any non-trivial statistical inference.

Access to the full statistical model, that is, the model before profiling or marginalization, is of enormous benefit. The full model makes it possible to test new signal hypotheses, while correctly accounting for non-Gaussian effects and nuisance parameter-induced correlations. Through appropriate updates to the full statistical model, analyses can be updated if more precise theoretical calculations, updated parton distribution functions (PDFs) and/or improved experimental calibrations become available. 
Finally, detailed information about nuisance parameters facilitates the combination of analyses and global fits.
Ideally, naming conventions of nuisance parameters should be standardised among analyses to this end.

The release of pyhf~\cite{Heinrich:2021gyp} and the first full HistFactory models by the ATLAS collaboration~\cite{ATL-PHYS-PUB-2019-029} in 2019 was met with enthusiasm by the phenomenology community. All major public reinterpretation tools quickly developed interfaces to pyhf. A recent search for statistical models on HEPData or on the ATLAS public results webpage~\cite{atlas:publicresults} resulted in 41 hits, of which 16 are from searches for supersymmetry and 21 from top-quark analyses. The former statistical models are systematically integrated and used in BSM reinterpretation tools,\footnote{A search on HEPData for, e.g.,  ``\texttt{analysis:(histfactory AND smodels)}'' or ``\texttt{analysis:(histfactory AND checkmate)}'' shows that 14 of the 16 searches that provide a HistFactory model are being reused by these tools.} while the latter have been shown to be highly useful in the EFT context~\cite{Elmer:2023wtr}.     
The CMS collaboration has recently followed suit by publishing their Combine software~\cite{CMS:2024onh} together with the 2012 Higgs boson observation statistical model~\cite{cms_collaboration_2024_c2948-e8875}, and the collaboration plans to systematically release full statistical models~\cite{Sekmen:2025hwq}.

Statistical models released in the pyhf~\cite{Heinrich:2021gyp} framework have led to it quickly becoming a top-cited paper. This is consistent with the expectation~\cite{Cranmer:2021urp} that the release of full statistical models will have high impact in the short, medium and long term. We therefore strongly encourage experimental collaborations to make the publication of full statistical models their standard practice. The development and adoption of the HEP Statistics Serialization Standard (HS3)~\cite{hs3:github} will further facilitate deployment and reuse in the community.

\section{Conclusion}

The LHC data are a unique scientific resource, and their long-term reusability will be an important legacy of the current scientific community for the generations that follow. The public investment in experimental programs 
warrants a concerted community-wide effort to ensure
that this investment continues to provide scientific value decades from now. 
At present, the work to preserve HEP results 
relies on the goodwill of a relatively small number of dedicated individuals. This model is not sustainable and a strategic European-wide approach is needed.
Preserving scientific results in a manner that is {\it useful} does not come for
free, however. Dedicated resources are needed, including well-defined career paths and reward structures to attract the highly talented scientific and engineering staff that will be needed. 
This  requires strategic prioritisation by the European community and commensurate resource allocation.
Without a strategic approach, the full impact of the public investment in experimental HEP will not be fully realized, to the detriment of the broader European scientific enterprise.

Services such as the CERN Open Data Portal, Zenodo, and HEPData (all European funded) are a good starting point and are having impact, and we note
the efforts around HEP data preservation discussed in a related contribution~\cite{DPHEP:ESPPU26}.
A concentrated and coordinated effort across the field is needed to make preservation --- with reuse in mind! --- standard
scientific practice.
Finally, it should be noted that many of the arguments given in this paper, and the tools described, apply beyond the (HL-)LHC experiments, and that these data are a critical
resource for the design and eventual exploitation of future experiments that are currently under discussion.

\section*{Acknowledgements}

We thank the LHC Physics Center at CERN and the CERN Theory Department for support in the organisation of workshops associated to this work.

The editors and contributors to this document acknowledge partial support by the CHIST-ERA project OpenMAPP under grants ANR-23-CHRO-0006, EP/Y036360/1 and 2022/04/Y/ST2/00186; STFC under the SwiftHEP grant ST/Y005678/1; the US Department of Energy under grant DE-SC0010102;
the FAPESP grants no.\ 2018/25225-9 and 2021/01089-1; 
the SMASH COFUND project No.~101081355, co-funded by the Republic of Slovenia and the European Union from the European Regional Development Fund; 
the National Research Foundation of Korea, NRF, contracts NRF-2021R1I1A3048138, NRF-2018R1A6A1A06024970 and NRF-2008-00460; 
and the Excellence Cluster ORIGINS, funded by DFG under Germany’s Excellence Strategy – EXC 2094 – 390783311.


\clearpage

\begin{thebibliography}{100}

\bibitem{Kraml:2012sg}
S.~Kraml et~al., \emph{{Searches for New Physics: Les Houches Recommendations
  for the Presentation of LHC Results}},
  \href{https://doi.org/10.1140/epjc/s10052-012-1976-3}{\emph{Eur. Phys. J. C}
  {\bfseries 72} (2012) 1976}
  [\href{https://arxiv.org/abs/1203.2489}{{\ttfamily 1203.2489}}].

\bibitem{LHCReinterpretationForum:2020xtr}
{LHC Reinterpretation Forum}, \emph{{Reinterpretation of LHC Results for New
  Physics: Status and Recommendations after Run 2}},
  \href{https://doi.org/10.21468/SciPostPhys.9.2.022}{\emph{SciPost Phys.}
  {\bfseries 9} (2020) 022} [\href{https://arxiv.org/abs/2003.07868}{{\ttfamily
  2003.07868}}].

\bibitem{Cranmer:2021urp}
K.~Cranmer et~al., \emph{{Publishing statistical models: Getting the most out
  of particle physics experiments}},
  \href{https://doi.org/10.21468/SciPostPhys.12.1.037}{\emph{SciPost Phys.}
  {\bfseries 12} (2022) 037}
  [\href{https://arxiv.org/abs/2109.04981}{{\ttfamily 2109.04981}}].

\bibitem{Bailey:2022tdz}
S.~Bailey et~al., \emph{{Data and Analysis Preservation, Recasting, and
  Reinterpretation (Snowmass 2021)}},
  \href{https://arxiv.org/abs/2203.10057}{{\ttfamily 2203.10057}}.

\bibitem{Bailey:2022pdq}
S.~Bailey, K.S.~Cranmer, M.~Feickert, R.~Fine, S.~Kraml and C.~Lange,
  \emph{{Reinterpretation and Long-Term Preservation of Data and Code (Snowmass
  2021)}},  \href{https://arxiv.org/abs/2209.08054}{{\ttfamily 2209.08054}}.

\bibitem{Araz:2023mda}
J.Y.~Araz et~al., \emph{{Les Houches guide to reusable ML models in LHC
  analyses}},  \href{https://arxiv.org/abs/2312.14575}{{\ttfamily 2312.14575}}.

\bibitem{RIFtalks}
{ATLAS and CMS presentations at the workshops of the LHC Reinterpretation
  Forum, \url{https://indico.cern.ch/category/14156/}}, 2016--2025.

\bibitem{FAIR:2016}
M.~Wilkinson et~al., \emph{{The FAIR Guiding Principles for scientific data
  management and stewardship}},
  \href{https://doi.org/10.1038/sdata.2016.18}{\emph{Sci Data} {\bfseries 3}
  (2016) 160018}.

\bibitem{CERN-OPEN-DATA}
CERN, ``{C}{E}{R}{N} {O}pen {D}ata {P}ortal.'' \url{https://opendata.cern.ch/}.

\bibitem{CERN-OPEN-2020-013}
CERN, \emph{{CERN Open Data Policy for the LHC Experiments}},  Tech. Rep.
  \href{https://cds.cern.ch/record/2745133}{CERN-OPEN-2020-013}, CERN, Geneva
  (2020), \href{https://doi.org/10.17181/CERN.QXNK.8L2G}{DOI}.

\bibitem{mccauley_2025_15078671}
T.~McCauley, ``cms-dpoa/data-usage: 24-03-2025.''
  \href{https://doi.org/10.5281/zenodo.15078671}{{10.5281/zenodo.15078671}},
  Mar., 2025.

\bibitem{DPHEP:2023blx}
{\scshape DPHEP} collaboration, \emph{{Data preservation in high energy
  physics}}, \href{https://doi.org/10.1140/epjc/s10052-023-11885-1}{\emph{Eur.
  Phys. J. C} {\bfseries 83} (2023) 795}
  [\href{https://arxiv.org/abs/2302.03583}{{\ttfamily 2302.03583}}].

\bibitem{DPHEP:ESPPU26}
{\scshape DPHEP} collaboration, ``{Data Preservation in High Energy Physics}.''
  {document submitted to the ESPPU}, March, 2025.

\bibitem{Buckley:2019xhk}
A.~Buckley, P.~Ilten, D.~Konstantinov, L.~L\"onnblad, J.~Monk, W.~Pokorski
  et~al., \emph{{The HepMC3 event record library for Monte Carlo event
  generators}}, \href{https://doi.org/10.1016/j.cpc.2020.107310}{\emph{Comput.
  Phys. Commun.} {\bfseries 260} (2021) 107310}
  [\href{https://arxiv.org/abs/1912.08005}{{\ttfamily 1912.08005}}].

\bibitem{Marshall2025}
R.~Mahbubani and Z.~Marshall, ``Open event generation.'' Invited contribution
  to 9th workshop Reinterpretation of the LHC results for New Physics, 2025.

\bibitem{Maguire:2017ypu}
E.~Maguire, L.~Heinrich and G.~Watt, \emph{{HEPData: a repository for high
  energy physics data}},
  \href{https://doi.org/10.1088/1742-6596/898/10/102006}{\emph{J. Phys. Conf.
  Ser.} {\bfseries 898} (2017) 102006}
  [\href{https://arxiv.org/abs/1704.05473}{{\ttfamily 1704.05473}}].

\bibitem{Buckley:2010jn}
A.~Buckley and M.~Whalley, \emph{{HepData reloaded: Reinventing the HEP data
  archive}}, \href{https://doi.org/10.22323/1.093.0067}{\emph{PoS} {\bfseries
  ACAT2010} (2010) 067} [\href{https://arxiv.org/abs/1006.0517}{{\ttfamily
  1006.0517}}].

\bibitem{HEPData:notebook}
G.~Watt.
  {\url{https://github.com/HEPData/miscellaneous/blob/main/notebooks/count_inspire_records_with_hepdata.ipynb}}.

\bibitem{Kraml:2013mwa}
S.~Kraml, S.~Kulkarni, U.~Laa, A.~Lessa, W.~Magerl, D.~Proschofsky-Spindler
  et~al., \emph{{SModelS: a tool for interpreting simplified-model results from
  the LHC and its application to supersymmetry}},
  \href{https://doi.org/10.1140/epjc/s10052-014-2868-5}{\emph{Eur. Phys. J. C}
  {\bfseries 74} (2014) 2868}
  [\href{https://arxiv.org/abs/1312.4175}{{\ttfamily 1312.4175}}].

\bibitem{Ambrogi:2017neo}
F.~Ambrogi, S.~Kraml, S.~Kulkarni, U.~Laa, A.~Lessa, V.~Magerl et~al.,
  \emph{{SModelS v1.1 user manual: Improving simplified model constraints with
  efficiency maps}},
  \href{https://doi.org/10.1016/j.cpc.2018.02.007}{\emph{Comput. Phys. Commun.}
  {\bfseries 227} (2018) 72}
  [\href{https://arxiv.org/abs/1701.06586}{{\ttfamily 1701.06586}}].

\bibitem{Ambrogi:2018ujg}
F.~Ambrogi et~al., \emph{{SModelS v1.2: long-lived particles, combination of
  signal regions, and other novelties}},
  \href{https://doi.org/10.1016/j.cpc.2019.07.013}{\emph{Comput. Phys. Commun.}
  {\bfseries 251} (2020) 106848}
  [\href{https://arxiv.org/abs/1811.10624}{{\ttfamily 1811.10624}}].

\bibitem{Alguero:2021dig}
G.~Alguero, J.~Heisig, C.K.~Khosa, S.~Kraml, S.~Kulkarni, A.~Lessa et~al.,
  \emph{{Constraining new physics with SModelS version 2}},
  \href{https://doi.org/10.1007/JHEP08(2022)068}{\emph{JHEP} {\bfseries 08}
  (2022) 068} [\href{https://arxiv.org/abs/2112.00769}{{\ttfamily
  2112.00769}}].

\bibitem{MahdiAltakach:2023bdn}
M.M.~Altakach, S.~Kraml, A.~Lessa, S.~Narasimha, T.~Pascal and W.~Waltenberger,
  \emph{{SModelS v2.3: Enabling global likelihood analyses}},
  \href{https://doi.org/10.21468/SciPostPhys.15.5.185}{\emph{SciPost Phys.}
  {\bfseries 15} (2023) 185}
  [\href{https://arxiv.org/abs/2306.17676}{{\ttfamily 2306.17676}}].

\bibitem{Altakach:2024jwk}
M.M.~Altakach, S.~Kraml, A.~Lessa, S.~Narasimha, T.~Pascal, C.~Ramos et~al.,
  \emph{{SModelS v3: going beyond $ \mathcal{Z} _{2}$ topologies}},
  \href{https://doi.org/10.1007/JHEP11(2024)074}{\emph{JHEP} {\bfseries 11}
  (2024) 074} [\href{https://arxiv.org/abs/2409.12942}{{\ttfamily
  2409.12942}}].

\bibitem{Bechtle:2008jh}
P.~Bechtle, O.~Brein, S.~Heinemeyer, G.~Weiglein and K.E.~Williams,
  \emph{{HiggsBounds: Confronting Arbitrary Higgs Sectors with Exclusion Bounds
  from LEP and the Tevatron}},
  \href{https://doi.org/10.1016/j.cpc.2009.09.003}{\emph{Comput. Phys. Commun.}
  {\bfseries 181} (2010) 138}
  [\href{https://arxiv.org/abs/0811.4169}{{\ttfamily 0811.4169}}].

\bibitem{Bechtle:2011sb}
P.~Bechtle, O.~Brein, S.~Heinemeyer, G.~Weiglein and K.E.~Williams,
  \emph{{HiggsBounds 2.0.0: Confronting Neutral and Charged Higgs Sector
  Predictions with Exclusion Bounds from LEP and the Tevatron}},
  \href{https://doi.org/10.1016/j.cpc.2011.07.015}{\emph{Comput. Phys. Commun.}
  {\bfseries 182} (2011) 2605}
  [\href{https://arxiv.org/abs/1102.1898}{{\ttfamily 1102.1898}}].

\bibitem{Bechtle:2013xfa}
P.~Bechtle, S.~Heinemeyer, O.~St\r{a}l, T.~Stefaniak and G.~Weiglein,
  \emph{{$HiggsSignals$: Confronting arbitrary Higgs sectors with measurements
  at the Tevatron and the LHC}},
  \href{https://doi.org/10.1140/epjc/s10052-013-2711-4}{\emph{Eur. Phys. J. C}
  {\bfseries 74} (2014) 2711}
  [\href{https://arxiv.org/abs/1305.1933}{{\ttfamily 1305.1933}}].

\bibitem{Bechtle:2013wla}
P.~Bechtle, O.~Brein, S.~Heinemeyer, O.~St\r{a}l, T.~Stefaniak, G.~Weiglein
  et~al., \emph{{$\mathsf{HiggsBounds}-4$: Improved Tests of Extended Higgs
  Sectors against Exclusion Bounds from LEP, the Tevatron and the LHC}},
  \href{https://doi.org/10.1140/epjc/s10052-013-2693-2}{\emph{Eur. Phys. J. C}
  {\bfseries 74} (2014) 2693}
  [\href{https://arxiv.org/abs/1311.0055}{{\ttfamily 1311.0055}}].

\bibitem{Bechtle:2020pkv}
P.~Bechtle, D.~Dercks, S.~Heinemeyer, T.~Klingl, T.~Stefaniak, G.~Weiglein
  et~al., \emph{{HiggsBounds-5: Testing Higgs Sectors in the LHC 13 TeV Era}},
  \href{https://doi.org/10.1140/epjc/s10052-020-08557-9}{\emph{Eur. Phys. J. C}
  {\bfseries 80} (2020) 1211}
  [\href{https://arxiv.org/abs/2006.06007}{{\ttfamily 2006.06007}}].

\bibitem{Bechtle:2020uwn}
P.~Bechtle, S.~Heinemeyer, T.~Klingl, T.~Stefaniak, G.~Weiglein and
  J.~Wittbrodt, \emph{{HiggsSignals-2: Probing new physics with precision Higgs
  measurements in the LHC 13 TeV era}},
  \href{https://doi.org/10.1140/epjc/s10052-021-08942-y}{\emph{Eur. Phys. J. C}
  {\bfseries 81} (2021) 145}
  [\href{https://arxiv.org/abs/2012.09197}{{\ttfamily 2012.09197}}].

\bibitem{Bahl:2022igd}
H.~Bahl, T.~Biek\"otter, S.~Heinemeyer, C.~Li, S.~Paasch, G.~Weiglein et~al.,
  \emph{{HiggsTools: BSM scalar phenomenology with new versions of HiggsBounds
  and HiggsSignals}},
  \href{https://doi.org/10.1016/j.cpc.2023.108803}{\emph{Comput. Phys. Commun.}
  {\bfseries 291} (2023) 108803}
  [\href{https://arxiv.org/abs/2210.09332}{{\ttfamily 2210.09332}}].

\bibitem{Ilten:2018crw}
P.~Ilten, Y.~Soreq, M.~Williams and W.~Xue, \emph{{Serendipity in dark photon
  searches}}, \href{https://doi.org/10.1007/JHEP06(2018)004}{\emph{JHEP}
  {\bfseries 06} (2018) 004}
  [\href{https://arxiv.org/abs/1801.04847}{{\ttfamily 1801.04847}}].

\bibitem{Giani:2023gfq}
T.~Giani, G.~Magni and J.~Rojo, \emph{{SMEFiT: a flexible toolbox for global
  interpretations of particle physics data with effective field theories}},
  \href{https://doi.org/10.1140/epjc/s10052-023-11534-7}{\emph{Eur. Phys. J. C}
  {\bfseries 83} (2023) 393}
  [\href{https://arxiv.org/abs/2302.06660}{{\ttfamily 2302.06660}}].

\bibitem{Simko:2018zzz}
T.~\v{S}imko, L.~Heinrich, H.~Hirvonsalo, D.~Kousidis and D.~Rodr\'\i{}guez,
  \emph{{REANA: A System for Reusable Research Data Analyses}},
  \href{https://doi.org/10.1051/epjconf/201921406034}{\emph{EPJ Web Conf.}
  {\bfseries 214} (2019) 06034}.

\bibitem{ATLAS:2015wrn}
{\scshape {ATLAS}} collaboration, \emph{{Summary of the ATLAS
  experiment\textquoteright{}s sensitivity to supersymmetry after LHC Run 1
  \textemdash{} interpreted in the phenomenological MSSM}},
  \href{https://doi.org/10.1007/JHEP10(2015)134}{\emph{JHEP} {\bfseries 10}
  (2015) 134} [\href{https://arxiv.org/abs/1508.06608}{{\ttfamily
  1508.06608}}].

\bibitem{ATLAS:2016dei}
{\scshape {ATLAS}} collaboration, \emph{{Dark matter interpretations of ATLAS
  searches for the electroweak production of supersymmetric particles in $
  \sqrt{s}=8 $ TeV proton-proton collisions}},
  \href{https://doi.org/10.1007/JHEP09(2016)175}{\emph{JHEP} {\bfseries 09}
  (2016) 175} [\href{https://arxiv.org/abs/1608.00872}{{\ttfamily
  1608.00872}}].

\bibitem{ATLAS:2016hks}
{\scshape {ATLAS}} collaboration, \emph{{A re-interpretation of
  $\sqrt{s}=8~$TeV ATLAS results on electroweak supersymmetry production to
  explore general gauge mediated models}},  Tech. Rep.
  \href{https://cds.cern.ch/record/2198316}{ATLAS-CONF-2016-033} (7, 2016).

\bibitem{ATL-PHYS-PUB-2019-032}
{\scshape {ATLAS}} collaboration, \emph{{RECAST framework reinterpretation of
  an ATLAS Dark Matter Search constraining a model of a dark Higgs boson
  decaying to two $b$-quarks}},  Tech. Rep.
  \href{https://cds.cern.ch/record/2686290}{ATL-PHYS-PUB-2019-032}, CERN,
  Geneva (2019).

\bibitem{ATL-PHYS-PUB-2021-020}
{\scshape {ATLAS}} collaboration, \emph{{Constraining the Dark Sector with the
  monojet signature in the ATLAS experiment}},  Tech. Rep.
  \href{https://cds.cern.ch/record/2772627}{ATL-PHYS-PUB-2021-020}, CERN,
  Geneva (2021).

\bibitem{ATL-PHYS-PUB-2022-045}
{ATLAS collaboration}, \emph{{Active Learning reinterpretation of an ATLAS Dark
  Matter search constraining a model of a dark Higgs boson decaying to two
  b-quarks}},  Tech. Rep.
  \href{https://cds.cern.ch/record/2839789}{ATL-PHYS-PUB-2022-045}, CERN,
  Geneva (2022).

\bibitem{ATLAS:2024qmx}
{\scshape {ATLAS}} collaboration, \emph{{ATLAS Run 2 searches for electroweak
  production of supersymmetric particles interpreted within the pMSSM}},
  \href{https://doi.org/10.1007/JHEP05(2024)106}{\emph{JHEP} {\bfseries 05}
  (2024) 106} [\href{https://arxiv.org/abs/2402.01392}{{\ttfamily
  2402.01392}}].

\bibitem{Bierlich:2019rhm}
C.~Bierlich et~al., \emph{{Robust Independent Validation of Experiment and
  Theory: Rivet version 3}},
  \href{https://doi.org/10.21468/SciPostPhys.8.2.026}{\emph{SciPost Phys.}
  {\bfseries 8} (2020) 026} [\href{https://arxiv.org/abs/1912.05451}{{\ttfamily
  1912.05451}}].

\bibitem{Drees:2013wra}
M.~Drees, H.~Dreiner, D.~Schmeier, J.~Tattersall and J.S.~Kim,
  \emph{{CheckMATE: Confronting your Favourite New Physics Model with LHC
  Data}}, \href{https://doi.org/10.1016/j.cpc.2014.10.018}{\emph{Comput. Phys.
  Commun.} {\bfseries 187} (2015) 227}
  [\href{https://arxiv.org/abs/1312.2591}{{\ttfamily 1312.2591}}].

\bibitem{Dercks:2016npn}
D.~Dercks, N.~Desai, J.S.~Kim, K.~Rolbiecki, J.~Tattersall and T.~Weber,
  \emph{{CheckMATE 2: From the model to the limit}},
  \href{https://doi.org/10.1016/j.cpc.2017.08.021}{\emph{Comput. Phys. Commun.}
  {\bfseries 221} (2017) 383}
  [\href{https://arxiv.org/abs/1611.09856}{{\ttfamily 1611.09856}}].

\bibitem{Dumont:2014tja}
B.~Dumont, B.~Fuks, S.~Kraml, S.~Bein, G.~Chalons, E.~Conte et~al.,
  \emph{{Toward a public analysis database for LHC new physics searches using
  MADANALYSIS 5}},
  \href{https://doi.org/10.1140/epjc/s10052-014-3242-3}{\emph{Eur. Phys. J. C}
  {\bfseries 75} (2015) 56} [\href{https://arxiv.org/abs/1407.3278}{{\ttfamily
  1407.3278}}].

\bibitem{Conte:2018vmg}
E.~Conte and B.~Fuks, \emph{{Confronting new physics theories to LHC data with
  MADANALYSIS 5}}, \href{https://doi.org/10.1142/S0217751X18300272}{\emph{Int.
  J. Mod. Phys. A} {\bfseries 33} (2018) 1830027}
  [\href{https://arxiv.org/abs/1808.00480}{{\ttfamily 1808.00480}}].

\bibitem{Goodsell:2024aig}
M.D.~Goodsell, \emph{{HackAnalysis 2: A powerful and hackable recasting tool}},
   \href{https://arxiv.org/abs/2406.10042}{{\ttfamily 2406.10042}}.

\bibitem{GAMBIT:2017qxg}
{\scshape GAMBIT} collaboration, \emph{{ColliderBit: a GAMBIT module for the
  calculation of high-energy collider observables and likelihoods}},
  \href{https://doi.org/10.1140/epjc/s10052-017-5285-8}{\emph{Eur. Phys. J. C}
  {\bfseries 77} (2017) 795}
  [\href{https://arxiv.org/abs/1705.07919}{{\ttfamily 1705.07919}}].

\bibitem{Buckley:2019stt}
A.~Buckley, D.~Kar and K.~Nordstr\"om, \emph{{Fast simulation of detector
  effects in Rivet}},
  \href{https://doi.org/10.21468/SciPostPhys.8.2.025}{\emph{SciPost Phys.}
  {\bfseries 8} (2020) 025} [\href{https://arxiv.org/abs/1910.01637}{{\ttfamily
  1910.01637}}].

\bibitem{ATLAS:2022yru}
{\scshape {ATLAS}} collaboration, \emph{{SimpleAnalysis: Truth-level Analysis
  Framework}},  Tech. Rep.
  \href{https://cds.cern.ch/record/2805991}{ATL-PHYS-PUB-2022-017} (2022),
  \href{https://doi.org/10.17181/CERN.R6S3.0QKV}{DOI}.

\bibitem{simpleanalysis:github}
``{SimpleAnalysis, v1.1.0}.'' {\url{https://simpleanalysis.docs.cern.ch}}.

\bibitem{Prosper:2022lnf}
H.B.~Prosper, S.~Sekmen and G.~Unel, \emph{{Analysis Description Language: A
  DSL for HEP Analysis}},  in \emph{{Snowmass 2021}}, 3, 2022
  [\href{https://arxiv.org/abs/2203.09886}{{\ttfamily 2203.09886}}].

\bibitem{habedank_2025_15076716}
M.~Habedank, ``Code "reinterpretation metadata analysis".'' DOI:
  \href{https://doi.org/10.5281/zenodo.15076716}{10.5281/zenodo.15076716},
  Mar., 2025.

\bibitem{Conte:2014zja}
E.~Conte, B.~Dumont, B.~Fuks and C.~Wymant, \emph{{Designing and recasting LHC
  analyses with MadAnalysis 5}},
  \href{https://doi.org/10.1140/epjc/s10052-014-3103-0}{\emph{Eur. Phys. J. C}
  {\bfseries 74} (2014) 3103}
  [\href{https://arxiv.org/abs/1405.3982}{{\ttfamily 1405.3982}}].

\bibitem{GAMBIT:2017yxo}
{\scshape GAMBIT} collaboration, \emph{{GAMBIT: The Global and Modular
  Beyond-the-Standard-Model Inference Tool}},
  \href{https://doi.org/10.1140/epjc/s10052-017-5321-8}{\emph{Eur. Phys. J. C}
  {\bfseries 77} (2017) 784}
  [\href{https://arxiv.org/abs/1705.07908}{{\ttfamily 1705.07908}}].

\bibitem{Kvellestad:2019vxm}
A.~Kvellestad, P.~Scott and M.~White, \emph{{GAMBIT and its application in the
  search for physics Beyond the Standard Model}},
  \href{https://doi.org/10.1016/j.ppnp.2020.103769}{\emph{Prog. Part. Nucl.
  Phys.} {\bfseries 113} (2020) 103769}
  [\href{https://arxiv.org/abs/1912.04079}{{\ttfamily 1912.04079}}].

\bibitem{Alvarez:2020yim}
E.~Alvarez, M.~Est\'evez and R.M.~Sand\'a~Seoane, \emph{{Z'-explorer: A simple
  tool to probe Z' models against LHC data}},
  \href{https://doi.org/10.1016/j.cpc.2021.108144}{\emph{Comput. Phys. Commun.}
  {\bfseries 269} (2021) 108144}
  [\href{https://arxiv.org/abs/2005.05194}{{\ttfamily 2005.05194}}].

\bibitem{Buckley:2021neu}
A.~Buckley et~al., \emph{{Testing new physics models with global comparisons to
  collider measurements: the Contur toolkit}},
  \href{https://doi.org/10.21468/SciPostPhysCore.4.2.013}{\emph{SciPost Phys.
  Core} {\bfseries 4} (2021) 013}
  [\href{https://arxiv.org/abs/2102.04377}{{\ttfamily 2102.04377}}].

\bibitem{Desai:2021jsa}
{\scshape CheckMATE} collaboration, \emph{{Constraining electroweak and
  strongly charged long-lived particles with CheckMATE}},
  \href{https://doi.org/10.1140/epjc/s10052-021-09727-z}{\emph{Eur. Phys. J. C}
  {\bfseries 81} (2021) 968}
  [\href{https://arxiv.org/abs/2104.04542}{{\ttfamily 2104.04542}}].

\bibitem{Lozano:2021zbu}
V.M.~Lozano, R.M.S.~Seoane and J.~Zurita, \emph{{Z'-explorer 2.0:
  Reconnoitering the dark matter landscape}},
  \href{https://doi.org/10.1016/j.cpc.2023.108729}{\emph{Comput. Phys. Commun.}
  {\bfseries 288} (2023) 108729}
  [\href{https://arxiv.org/abs/2109.13194}{{\ttfamily 2109.13194}}].

\bibitem{Araz:2021akd}
J.Y.~Araz, B.~Fuks, M.D.~Goodsell and M.~Utsch, \emph{{Recasting LHC searches
  for long-lived particles with MadAnalysis~5}},
  \href{https://doi.org/10.1140/epjc/s10052-022-10511-w}{\emph{Eur. Phys. J. C}
  {\bfseries 82} (2022) 597}
  [\href{https://arxiv.org/abs/2112.05163}{{\ttfamily 2112.05163}}].

\bibitem{Bernon:2015hsa}
J.~Bernon and B.~Dumont, \emph{{Lilith: a tool for constraining new physics
  from Higgs measurements}},
  \href{https://doi.org/10.1140/epjc/s10052-015-3645-9}{\emph{Eur. Phys. J. C}
  {\bfseries 75} (2015) 440}
  [\href{https://arxiv.org/abs/1502.04138}{{\ttfamily 1502.04138}}].

\bibitem{Kraml:2019sis}
S.~Kraml, T.Q.~Loc, D.T.~Nhung and L.D.~Ninh, \emph{{Constraining new physics
  from Higgs measurements with Lilith: update to LHC Run 2 results}},
  \href{https://doi.org/10.21468/SciPostPhys.7.4.052}{\emph{SciPost Phys.}
  {\bfseries 7} (2019) 052} [\href{https://arxiv.org/abs/1908.03952}{{\ttfamily
  1908.03952}}].

\bibitem{Buckley:2010ar}
A.~Buckley, J.~Butterworth, D.~Grellscheid, H.~Hoeth, L.~Lonnblad, J.~Monk
  et~al., \emph{{Rivet user manual}},
  \href{https://doi.org/10.1016/j.cpc.2013.05.021}{\emph{Comput. Phys. Commun.}
  {\bfseries 184} (2013) 2803}
  [\href{https://arxiv.org/abs/1003.0694}{{\ttfamily 1003.0694}}].

\bibitem{Bierlich:2020wms}
C.~Bierlich et~al., \emph{{Confronting experimental data with heavy-ion models:
  RIVET for heavy ions}},
  \href{https://doi.org/10.1140/epjc/s10052-020-8033-4}{\emph{Eur. Phys. J. C}
  {\bfseries 80} (2020) 485}
  [\href{https://arxiv.org/abs/2001.10737}{{\ttfamily 2001.10737}}].

\bibitem{Bierlich:2024vqo}
C.~Bierlich, A.~Buckley, J.M.~Butterworth, C.~Gutschow, L.~Lonnblad, T.~Procter
  et~al., \emph{{Robust independent validation of experiment and theory: Rivet
  version 4 release note}},
  \href{https://doi.org/10.21468/SciPostPhysCodeb.36}{\emph{SciPost Phys.
  Codeb.} {\bfseries 36} (2024) 1}
  [\href{https://arxiv.org/abs/2404.15984}{{\ttfamily 2404.15984}}].

\bibitem{LaCagnina:2023yvi}
S.~La~Cagnina, K.~Kr\"oninger, S.~Kluth and A.~Verbytskyi, \emph{{A Bayesian
  tune of the Herwig Monte Carlo event generator}},
  \href{https://doi.org/10.1088/1748-0221/18/10/P10033}{\emph{JINST} {\bfseries
  18} (2023) P10033} [\href{https://arxiv.org/abs/2302.01139}{{\ttfamily
  2302.01139}}].

\bibitem{Baum:2023inl}
S.~Baum, M.~Carena, T.~Ou, D.~Rocha, N.R.~Shah and C.E.M.~Wagner,
  \emph{{Lighting up the LHC with Dark Matter}},
  \href{https://doi.org/10.1007/JHEP11(2023)037}{\emph{JHEP} {\bfseries 11}
  (2023) 037} [\href{https://arxiv.org/abs/2303.01523}{{\ttfamily
  2303.01523}}].

\bibitem{Araz:2023axv}
J.Y.~Araz, A.~Buckley and B.~Fuks, \emph{{Searches for new physics with boosted
  top quarks in the MadAnalysis 5 and Rivet frameworks}},
  \href{https://doi.org/10.1140/epjc/s10052-023-11779-2}{\emph{Eur. Phys. J. C}
  {\bfseries 83} (2023) 664}
  [\href{https://arxiv.org/abs/2303.03427}{{\ttfamily 2303.03427}}].

\bibitem{Chakraborty:2023hrk}
A.~Chakraborty, S.~Dasmahapatra, H.~Day-Hall, B.~Ford, S.~Jain and S.~Moretti,
  \emph{{Fat b-jet analyses using old and new clustering algorithms in new
  Higgs boson searches at the LHC}},
  \href{https://doi.org/10.1140/epjc/s10052-023-11537-4}{\emph{Eur. Phys. J. C}
  {\bfseries 83} (2023) 347}
  [\href{https://arxiv.org/abs/2303.05189}{{\ttfamily 2303.05189}}].

\bibitem{Chang:2023cki}
C.~Chang, P.~Scott, T.E.~Gonzalo, F.~Kahlhoefer and M.~White, \emph{{Global
  fits of simplified models for dark matter with GAMBIT: II. Vector dark matter
  with an s-channel vector mediator}},
  \href{https://doi.org/10.1140/epjc/s10052-023-11859-3}{\emph{Eur. Phys. J. C}
  {\bfseries 83} (2023) 692}
  [\href{https://arxiv.org/abs/2303.08351}{{\ttfamily 2303.08351}}].

\bibitem{GAMBIT:2023yih}
{\scshape GAMBIT} collaboration, \emph{{Collider constraints on electroweakinos
  in the presence of a light gravitino}},
  \href{https://doi.org/10.1140/epjc/s10052-023-11574-z}{\emph{Eur. Phys. J. C}
  {\bfseries 83} (2023) 493}
  [\href{https://arxiv.org/abs/2303.09082}{{\ttfamily 2303.09082}}].

\bibitem{Andersen:2023cku}
J.R.~Andersen, A.~Maier and D.~Ma\^\i{}tre, \emph{{Efficient negative-weight
  elimination in large high-multiplicity Monte Carlo event samples}},
  \href{https://doi.org/10.1140/epjc/s10052-023-11905-0}{\emph{Eur. Phys. J. C}
  {\bfseries 83} (2023) 835}
  [\href{https://arxiv.org/abs/2303.15246}{{\ttfamily 2303.15246}}].

\bibitem{Dreiner:2023bvs}
H.K.~Dreiner, Y.S.~Koay, D.~K\"ohler, V.M.~Lozano, J.~Montejo~Berlingen,
  S.~Nangia et~al., \emph{{The ABC of RPV: classification of R-parity violating
  signatures at the LHC for small couplings}},
  \href{https://doi.org/10.1007/JHEP07(2023)215}{\emph{JHEP} {\bfseries 07}
  (2023) 215} [\href{https://arxiv.org/abs/2306.07317}{{\ttfamily
  2306.07317}}].

\bibitem{Carrasco:2023loy}
J.~Carrasco and J.~Zurita, \emph{{Emerging jet probes of strongly interacting
  dark sectors}}, \href{https://doi.org/10.1007/JHEP01(2024)034}{\emph{JHEP}
  {\bfseries 01} (2024) 034}
  [\href{https://arxiv.org/abs/2307.04847}{{\ttfamily 2307.04847}}].

\bibitem{Araz:2023bwx}
J.Y.~Araz, \emph{{Spey: Smooth inference for reinterpretation studies}},
  \href{https://doi.org/10.21468/SciPostPhys.16.1.032}{\emph{SciPost Phys.}
  {\bfseries 16} (2024) 032}
  [\href{https://arxiv.org/abs/2307.06996}{{\ttfamily 2307.06996}}].

\bibitem{Baer:2023uwo}
H.~Baer, V.~Barger, J.~Dutta, D.~Sengupta and K.~Zhang, \emph{{Top squarks from
  the landscape at high luminosity LHC}},
  \href{https://doi.org/10.1103/PhysRevD.108.075027}{\emph{Phys. Rev. D}
  {\bfseries 108} (2023) 075027}
  [\href{https://arxiv.org/abs/2307.08067}{{\ttfamily 2307.08067}}].

\bibitem{Arina:2023msd}
C.~Arina, B.~Fuks, J.~Heisig, M.~Kr\"amer, L.~Mantani and L.~Panizzi,
  \emph{{Comprehensive exploration of t-channel simplified models of dark
  matter}}, \href{https://doi.org/10.1103/PhysRevD.108.115007}{\emph{Phys. Rev.
  D} {\bfseries 108} (2023) 115007}
  [\href{https://arxiv.org/abs/2307.10367}{{\ttfamily 2307.10367}}].

\bibitem{Bahl:2023xkw}
H.~Bahl, S.~Koren and L.-T.~Wang, \emph{{Topportunities at the LHC: rare top
  decays with light singlets}},
  \href{https://doi.org/10.1140/epjc/s10052-024-13411-3}{\emph{Eur. Phys. J. C}
  {\bfseries 84} (2024) 1100}
  [\href{https://arxiv.org/abs/2307.11154}{{\ttfamily 2307.11154}}].

\bibitem{Brahma:2023psr}
N.~Brahma, S.~Heeba and K.~Schutz, \emph{{Resonant pseudo-Dirac dark matter as
  a sub-GeV thermal target}},
  \href{https://doi.org/10.1103/PhysRevD.109.035006}{\emph{Phys. Rev. D}
  {\bfseries 109} (2024) 035006}
  [\href{https://arxiv.org/abs/2308.01960}{{\ttfamily 2308.01960}}].

\bibitem{FASER:2023tle}
{\scshape FASER} collaboration, \emph{{Search for dark photons with the FASER
  detector at the LHC}},
  \href{https://doi.org/10.1016/j.physletb.2023.138378}{\emph{Phys. Lett. B}
  {\bfseries 848} (2024) 138378}
  [\href{https://arxiv.org/abs/2308.05587}{{\ttfamily 2308.05587}}].

\bibitem{KA:2023dyz}
S.~K.~A., A.~Das, G.~Lambiase, T.~Nomura and Y.~Orikasa, \emph{{Probing chiral
  and flavored $Z^\prime $ from cosmic bursts through neutrino interactions}},
  \href{https://doi.org/10.1140/epjc/s10052-024-13530-x}{\emph{Eur. Phys. J. C}
  {\bfseries 84} (2024) 1224}
  [\href{https://arxiv.org/abs/2308.14483}{{\ttfamily 2308.14483}}].

\bibitem{Coloma:2023oxx}
P.~Coloma, J.~Mart\'\i{}n-Albo and S.~Urrea, \emph{{Discovering long-lived
  particles at DUNE}},
  \href{https://doi.org/10.1103/PhysRevD.109.035013}{\emph{Phys. Rev. D}
  {\bfseries 109} (2024) 035013}
  [\href{https://arxiv.org/abs/2309.06492}{{\ttfamily 2309.06492}}].

\bibitem{Carpenter:2023agq}
L.M.~Carpenter, H.~Gilmer, J.~Kawamura and T.~Murphy, \emph{{Taking aim at the
  wino-Higgsino plane with the LHC}},
  \href{https://doi.org/10.1103/PhysRevD.109.015012}{\emph{Phys. Rev. D}
  {\bfseries 109} (2024) 015012}
  [\href{https://arxiv.org/abs/2309.07213}{{\ttfamily 2309.07213}}].

\bibitem{Fieg:2023kld}
M.~Fieg, F.~Kling, H.~Schulz and T.~Sj\"ostrand, \emph{{Tuning pythia for
  forward physics experiments}},
  \href{https://doi.org/10.1103/PhysRevD.109.016010}{\emph{Phys. Rev. D}
  {\bfseries 109} (2024) 016010}
  [\href{https://arxiv.org/abs/2309.08604}{{\ttfamily 2309.08604}}].

\bibitem{Butterworth:2023rnw}
J.~Butterworth, H.~Debnath, P.~Fileviez~Perez and F.~Mitchell, \emph{{Custodial
  symmetry breaking and Higgs boson signatures at the LHC}},
  \href{https://doi.org/10.1103/PhysRevD.109.095014}{\emph{Phys. Rev. D}
  {\bfseries 109} (2024) 095014}
  [\href{https://arxiv.org/abs/2309.10027}{{\ttfamily 2309.10027}}].

\bibitem{Acaroglu:2023cza}
H.~Acaro\u{g}lu, M.~Blanke and M.~Tabet, \emph{{Opening the Higgs portal to
  lepton-flavoured dark matter}},
  \href{https://doi.org/10.1007/JHEP11(2023)079}{\emph{JHEP} {\bfseries 11}
  (2023) 079} [\href{https://arxiv.org/abs/2309.10700}{{\ttfamily
  2309.10700}}].

\bibitem{Buonocore:2023kna}
L.~Buonocore, F.~Kling, L.~Rottoli and J.~Sominka, \emph{{Predictions for
  neutrinos and new physics from forward heavy hadron production at the LHC}},
  \href{https://doi.org/10.1140/epjc/s10052-024-12726-5}{\emph{Eur. Phys. J. C}
  {\bfseries 84} (2024) 363}
  [\href{https://arxiv.org/abs/2309.12793}{{\ttfamily 2309.12793}}].

\bibitem{Dey:2023exa}
A.~Dey, V.~Keus, S.~Moretti and C.~Shepherd-Themistocleous, \emph{{A smoking
  gun signature of the 3HDM}},
  \href{https://doi.org/10.1007/JHEP07(2024)038}{\emph{JHEP} {\bfseries 07}
  (2024) 038} [\href{https://arxiv.org/abs/2310.06593}{{\ttfamily
  2310.06593}}].

\bibitem{Agin:2023yoq}
D.~Agin, B.~Fuks, M.D.~Goodsell and T.~Murphy, \emph{{Monojets reveal
  overlapping excesses for light compressed higgsinos}},
  \href{https://doi.org/10.1016/j.physletb.2024.138597}{\emph{Phys. Lett. B}
  {\bfseries 853} (2024) 138597}
  [\href{https://arxiv.org/abs/2311.17149}{{\ttfamily 2311.17149}}].

\bibitem{Baer:2023ech}
H.~Baer, V.~Barger, J.~Bolich, J.~Dutta and D.~Sengupta, \emph{{Natural anomaly
  mediation from the landscape with implications for LHC SUSY searches}},
  \href{https://doi.org/10.1103/PhysRevD.109.035011}{\emph{Phys. Rev. D}
  {\bfseries 109} (2024) 035011}
  [\href{https://arxiv.org/abs/2311.18120}{{\ttfamily 2311.18120}}].

\bibitem{Banerjee:2023upj}
A.~Banerjee, V.~Ellajosyula and L.~Panizzi, \emph{{Heavy vector-like quarks
  decaying to exotic scalars: a case study with triplets}},
  \href{https://doi.org/10.1007/JHEP01(2024)187}{\emph{JHEP} {\bfseries 01}
  (2024) 187} [\href{https://arxiv.org/abs/2311.17877}{{\ttfamily
  2311.17877}}].

\bibitem{Lessa:2023tqc}
A.~Lessa and V.~Sanz, \emph{{Going beyond Top EFT}},
  \href{https://doi.org/10.1007/JHEP04(2024)107}{\emph{JHEP} {\bfseries 04}
  (2024) 107} [\href{https://arxiv.org/abs/2312.00670}{{\ttfamily
  2312.00670}}].

\bibitem{Frank:2023epx}
M.~Frank, B.~Fuks, A.~Jueid, S.~Moretti and O.~Ozdal, \emph{{A novel search
  strategy for right-handed charged gauge bosons at the Large Hadron
  Collider}}, \href{https://doi.org/10.1007/JHEP02(2024)150}{\emph{JHEP}
  {\bfseries 02} (2024) 150}
  [\href{https://arxiv.org/abs/2312.08521}{{\ttfamily 2312.08521}}].

\bibitem{Acaroglu:2023phy}
H.~Acaro\u{g}lu, M.~Blanke, J.~Heisig, M.~Kr\"amer and L.~Rathmann,
  \emph{{Flavoured Majorana Dark Matter then and now: from freeze-out scenarios
  to LHC signatures}},
  \href{https://doi.org/10.1007/JHEP06(2024)179}{\emph{JHEP} {\bfseries 06}
  (2024) 179} [\href{https://arxiv.org/abs/2312.09274}{{\ttfamily
  2312.09274}}].

\bibitem{Carpenter:2023aec}
L.M.~Carpenter, K.~Schwind and T.~Murphy, \emph{{Leptonic signatures of
  color-sextet scalars. II. Exploiting unique large-ETmiss signals at the
  LHC}}, \href{https://doi.org/10.1103/PhysRevD.109.075010}{\emph{Phys. Rev. D}
  {\bfseries 109} (2024) 075010}
  [\href{https://arxiv.org/abs/2312.09273}{{\ttfamily 2312.09273}}].

\bibitem{Bagnaschi:2023cxg}
E.~Bagnaschi, G.~Corcella, R.~Franceschini and D.~Sengupta, \emph{{Rise and
  Fall of Light Top Squarks in the LHC Top-Quark Sample}},
  \href{https://doi.org/10.1103/PhysRevLett.133.061801}{\emph{Phys. Rev. Lett.}
  {\bfseries 133} (2024) 061801}
  [\href{https://arxiv.org/abs/2312.09794}{{\ttfamily 2312.09794}}].

\bibitem{Amram:2023jlc}
D.~Amram, K.~Bouzoud, N.~Chanon, H.~Hansen, M.R.~Ribeiro, Jr. and M.~Schreck,
  \emph{{New Constraint for Isotropic Lorentz Violation from LHC Data}},
  \href{https://doi.org/10.1103/PhysRevLett.132.211801}{\emph{Phys. Rev. Lett.}
  {\bfseries 132} (2024) 211801}
  [\href{https://arxiv.org/abs/2312.11307}{{\ttfamily 2312.11307}}].

\bibitem{Altakach:2023tsd}
M.M.~Altakach, S.~Kraml, A.~Lessa, S.~Narasimha, T.~Pascal, T.~Reymermier
  et~al., \emph{{Global LHC constraints on electroweak-inos with SModelS
  v2.3}}, \href{https://doi.org/10.21468/SciPostPhys.16.4.101}{\emph{SciPost
  Phys.} {\bfseries 16} (2024) 101}
  [\href{https://arxiv.org/abs/2312.16635}{{\ttfamily 2312.16635}}].

\bibitem{DeRomeri:2024dbv}
V.~De~Romeri, D.K.~Papoulias and C.A.~Ternes, \emph{{Light vector mediators at
  direct detection experiments}},
  \href{https://doi.org/10.1007/JHEP05(2024)165}{\emph{JHEP} {\bfseries 05}
  (2024) 165} [\href{https://arxiv.org/abs/2402.05506}{{\ttfamily
  2402.05506}}].

\bibitem{Barman:2024xlc}
R.K.~Barman, G.~B\'elanger, B.~Bhattacherjee, R.~Godbole and R.~Sengupta,
  \emph{{Current status of the light neutralino thermal dark matter in the
  phenomenological MSSM}},
  \href{https://doi.org/10.1103/PhysRevD.111.015014}{\emph{Phys. Rev. D}
  {\bfseries 111} (2025) 015014}
  [\href{https://arxiv.org/abs/2402.07991}{{\ttfamily 2402.07991}}].

\bibitem{Gartner:2024muk}
L.~G\"artner, N.~Hartmann, L.~Heinrich, M.~Horstmann, T.~Kuhr, M.~Reboud
  et~al., \emph{{Constructing model-agnostic likelihoods, a method for the
  reinterpretation of particle physics results}},
  \href{https://doi.org/10.1140/epjc/s10052-024-13038-4}{\emph{Eur. Phys. J. C}
  {\bfseries 84} (2024) 693}
  [\href{https://arxiv.org/abs/2402.08417}{{\ttfamily 2402.08417}}].

\bibitem{Kalinowski:2024uxe}
J.~Kalinowski and W.~Kotlarski, \emph{{Interpreting 95 GeV di-photon/$
  b\overline{b} $ excesses as a lightest Higgs boson of the MRSSM}},
  \href{https://doi.org/10.1007/JHEP07(2024)037}{\emph{JHEP} {\bfseries 07}
  (2024) 037} [\href{https://arxiv.org/abs/2403.08720}{{\ttfamily
  2403.08720}}].

\bibitem{Dehghani:2024lgz}
P.~Dehghani, M.~Frank and B.~Fuks, \emph{{Collider imprint of vector-like
  leptons in light of anomalous magnetic moment and neutrino data}},
  \href{https://doi.org/10.1140/epjc/s10052-024-13659-9}{\emph{Eur. Phys. J. C}
  {\bfseries 84} (2024) 1284}
  [\href{https://arxiv.org/abs/2403.11862}{{\ttfamily 2403.11862}}].

\bibitem{Feike:2024zfz}
A.~Feike, J.~Fiaschi, B.~Fuks, M.~Klasen and A.~Puck~Neuwirth,
  \emph{{Combination and reinterpretation of LHC SUSY searches}},
  \href{https://doi.org/10.1007/JHEP07(2024)122}{\emph{JHEP} {\bfseries 07}
  (2024) 122} [\href{https://arxiv.org/abs/2403.11715}{{\ttfamily
  2403.11715}}].

\bibitem{Chakraborti:2024pdn}
M.~Chakraborti, S.~Heinemeyer and I.~Saha, \emph{{Consistent excesses in the
  search for ${\tilde{\chi }_{2}^0\tilde{\chi }_{1}^\pm :}$ wino/bino vs.
  Higgsino dark matter}},
  \href{https://doi.org/10.1140/epjc/s10052-024-13180-z}{\emph{Eur. Phys. J. C}
  {\bfseries 84} (2024) 812}
  [\href{https://arxiv.org/abs/2403.14759}{{\ttfamily 2403.14759}}].

\bibitem{Cacciapaglia:2024wdn}
G.~Cacciapaglia, A.~Deandrea, M.~Kunkel and W.~Porod, \emph{{Coloured spin-1
  states in composite Higgs models}},
  \href{https://doi.org/10.1007/JHEP06(2024)092}{\emph{JHEP} {\bfseries 06}
  (2024) 092} [\href{https://arxiv.org/abs/2404.02198}{{\ttfamily
  2404.02198}}].

\bibitem{Baruah:2024gwy}
R.~Baruah, S.~Mondal, S.K.~Patra and S.~Roy, \emph{{Probing intractable
  beyond-standard-model parameter spaces armed with machine learning}},
  \href{https://doi.org/10.1140/epjs/s11734-024-01236-w}{\emph{Eur. Phys. J.
  ST} {\bfseries 233} (2024) 2597}
  [\href{https://arxiv.org/abs/2404.02698}{{\ttfamily 2404.02698}}].

\bibitem{Argyropoulos:2024yxo}
S.~Argyropoulos, U.~Haisch and I.~Kalaitzidou, \emph{{Novel collider signatures
  in the type-I 2HDM+a model}},
  \href{https://doi.org/10.1007/JHEP07(2024)263}{\emph{JHEP} {\bfseries 07}
  (2024) 263} [\href{https://arxiv.org/abs/2404.05704}{{\ttfamily
  2404.05704}}].

\bibitem{Agin:2024yfs}
D.~Agin, B.~Fuks, M.D.~Goodsell and T.~Murphy, \emph{{Seeking a coherent
  explanation of LHC excesses for compressed spectra}},
  \href{https://doi.org/10.1140/epjc/s10052-024-13594-9}{\emph{Eur. Phys. J. C}
  {\bfseries 84} (2024) 1218}
  [\href{https://arxiv.org/abs/2404.12423}{{\ttfamily 2404.12423}}].

\bibitem{Heisig:2024xbh}
J.~Heisig, A.~Lessa and L.M.D.~Ramos, \emph{{Probing conversion-driven
  freeze-out at the LHC}},
  \href{https://doi.org/10.1103/PhysRevD.110.015031}{\emph{Phys. Rev. D}
  {\bfseries 110} (2024) 015031}
  [\href{https://arxiv.org/abs/2404.16086}{{\ttfamily 2404.16086}}].

\bibitem{Ellwanger:2024vvs}
U.~Ellwanger, C.~Hugonie, S.F.~King and S.~Moretti, \emph{{NMSSM explanation
  for excesses in the search for neutralinos and charginos and a 95 GeV Higgs
  boson}}, \href{https://doi.org/10.1140/epjc/s10052-024-13129-2}{\emph{Eur.
  Phys. J. C} {\bfseries 84} (2024) 788}
  [\href{https://arxiv.org/abs/2404.19338}{{\ttfamily 2404.19338}}].

\bibitem{Butterworth:2024eyr}
J.~Butterworth, H.~Debnath, P.~Fileviez~Perez and Y.~Yeh, \emph{{Dark matter
  from anomaly cancellation at the LHC}},
  \href{https://doi.org/10.1103/PhysRevD.110.075001}{\emph{Phys. Rev. D}
  {\bfseries 110} (2024) 075001}
  [\href{https://arxiv.org/abs/2405.03749}{{\ttfamily 2405.03749}}].

\bibitem{Balan:2024cmq}
S.~Balan et~al., \emph{{Resonant or asymmetric: the status of sub-GeV dark
  matter}}, \href{https://doi.org/10.1088/1475-7516/2025/01/053}{\emph{JCAP}
  {\bfseries 01} (2025) 053}
  [\href{https://arxiv.org/abs/2405.17548}{{\ttfamily 2405.17548}}].

\bibitem{Cruz-Martinez:2024cbz}
J.~Cruz-Martinez, S.~Forte, N.~Laurenti, T.R.~Rabemananjara and J.~Rojo,
  \emph{{LO, NLO, and NNLO parton distributions for LHC event generators}},
  \href{https://doi.org/10.1007/JHEP09(2024)088}{\emph{JHEP} {\bfseries 09}
  (2024) 088} [\href{https://arxiv.org/abs/2406.12961}{{\ttfamily
  2406.12961}}].

\bibitem{Boto:2024tzp}
R.~Boto, P.N.~Figueiredo, J.C.~Rom\~ao and J.a.P.~Silva, \emph{{Novel two
  component dark matter features in the Z$_{2}$ \texttimes{} Z$_{2}$ 3HDM}},
  \href{https://doi.org/10.1007/JHEP11(2024)108}{\emph{JHEP} {\bfseries 11}
  (2024) 108} [\href{https://arxiv.org/abs/2407.15933}{{\ttfamily
  2407.15933}}].

\bibitem{Corpe:2024ntq}
L.~Corpe, A.~Goudelis, S.~Jeannot and S.H.~Jeon, \emph{{Probing exotic
  long-lived particles from the prompt side using the CONTUR method}},
  \href{https://doi.org/10.1007/JHEP02(2025)033}{\emph{JHEP} {\bfseries 02}
  (2025) 033} [\href{https://arxiv.org/abs/2407.18710}{{\ttfamily
  2407.18710}}].

\bibitem{Arbelaez:2024lcr}
C.~Arbel\'aez, G.~Cottin, J.C.~Helo, M.~Hirsch and T.B.~de~Melo,
  \emph{{Long-lived particle phenomenology in one-loop neutrino mass models
  with dark matter}},
  \href{https://doi.org/10.1007/JHEP02(2025)049}{\emph{JHEP} {\bfseries 02}
  (2025) 049} [\href{https://arxiv.org/abs/2408.03364}{{\ttfamily
  2408.03364}}].

\bibitem{Butterworth:2024hvb}
J.M.~Butterworth, I.~Helenius, J.J.J.~Castella, B.~Pattengale, S.~Sanjrani and
  M.~Wing, \emph{{Modelling the underlying event in photon-initiated
  processes}},
  \href{https://doi.org/10.21468/SciPostPhys.17.6.158}{\emph{SciPost Phys.}
  {\bfseries 17} (2024) 158}
  [\href{https://arxiv.org/abs/2408.15842}{{\ttfamily 2408.15842}}].

\bibitem{Fuks:2024qdt}
B.~Fuks, M.D.~Goodsell and T.~Murphy, \emph{{Monojets from compressed weak
  frustrated dark matter}},
  \href{https://doi.org/10.1103/PhysRevD.111.055010}{\emph{Phys. Rev. D}
  {\bfseries 111} (2025) 055010}
  [\href{https://arxiv.org/abs/2409.03014}{{\ttfamily 2409.03014}}].

\bibitem{Fan:2024wvo}
Y.-Z.~Fan, Y.-Y.~Li, C.-T.~Lu, X.-Y.~Luo, T.-P.~Tang, V.Q.~Tran et~al.,
  \emph{{Current status of inert Higgs dark matter with dark fermions}},
  \href{https://doi.org/10.1103/PhysRevD.111.015046}{\emph{Phys. Rev. D}
  {\bfseries 111} (2025) 015046}
  [\href{https://arxiv.org/abs/2410.00638}{{\ttfamily 2410.00638}}].

\bibitem{Athron:2024rir}
P.~Athron, A.~Crivellin, T.E.~Gonzalo, S.~Iguro and C.~Sierra, \emph{{Global
  fit to the 2HDM with generic sources of flavour violation using GAMBIT}},
  \href{https://doi.org/10.1007/JHEP11(2024)133}{\emph{JHEP} {\bfseries 11}
  (2024) 133} [\href{https://arxiv.org/abs/2410.10493}{{\ttfamily
  2410.10493}}].

\bibitem{Lu:2024ade}
C.-T.~Lu, X.~Wang, X.~Wei and Y.~Wu, \emph{{Probing long-lived doubly charged
  scalar in the Georgi-Machacek model at the LHC and in far detectors}},
  \href{https://doi.org/10.1007/JHEP02(2025)149}{\emph{JHEP} {\bfseries 02}
  (2025) 149} [\href{https://arxiv.org/abs/2410.19561}{{\ttfamily
  2410.19561}}].

\bibitem{Cornell:2024dki}
A.S.~Cornell, B.~Fuks, M.D.~Goodsell and A.M.~Ncube, \emph{{Improving smuon
  searches with neural networks}},
  \href{https://doi.org/10.1140/epjc/s10052-025-13748-3}{\emph{Eur. Phys. J. C}
  {\bfseries 85} (2025) 51} [\href{https://arxiv.org/abs/2411.04526}{{\ttfamily
  2411.04526}}].

\bibitem{Kim:2014eva}
J.S.~Kim, K.~Rolbiecki, K.~Sakurai and J.~Tattersall, \emph{{'Stop' that
  ambulance! New physics at the LHC?}},
  \href{https://doi.org/10.1007/JHEP12(2014)010}{\emph{JHEP} {\bfseries 12}
  (2014) 010} [\href{https://arxiv.org/abs/1406.0858}{{\ttfamily 1406.0858}}].

\bibitem{Giudice:2022bpq}
G.F.~Giudice, M.~McCullough and D.~Teresi, \emph{{dE/dx from boosted long-lived
  particles}}, \href{https://doi.org/10.1007/JHEP08(2022)012}{\emph{JHEP}
  {\bfseries 08} (2022) 012}
  [\href{https://arxiv.org/abs/2205.04473}{{\ttfamily 2205.04473}}].

\bibitem{kraml_2025_15084457}
S.~Kraml, ``{Citation counts for reinterpretation tools}.'' jupyter notebook
  DOI: \href{https://doi.org/10.5281/zenodo.15084457}{10.5281/zenodo.15084457},
  Mar., 2025.

\bibitem{alice-o2}
{ALICE collaboration}. \url{https://github.com/AliceO2Group/AliceO2}.

\bibitem{atlas-athena}
{ATLAS collaboration}. \url{https://gitlab.cern.ch/atlas/athena}.

\bibitem{cms-sw}
{CMS collaboration}. \url{https://github.com/cms-sw/cmssw}.

\bibitem{lhcb-gaudi}
{LHCb collaboration}. \url{https://gitlab.cern.ch/lhcb/Gaudi}.

\bibitem{deFavereau:2013fsa}
{\scshape DELPHES 3} collaboration, \emph{{DELPHES 3, A modular framework for
  fast simulation of a generic collider experiment}},
  \href{https://doi.org/10.1007/JHEP02(2014)057}{\emph{JHEP} {\bfseries 02}
  (2014) 057} [\href{https://arxiv.org/abs/1307.6346}{{\ttfamily 1307.6346}}].

\bibitem{ATLAS:2024ocv}
{\scshape ATLAS} collaboration, \emph{{Search for neutral long-lived particles
  that decay into displaced jets in the ATLAS calorimeter in association with
  leptons or jets using pp collisions at $ \sqrt{\textrm{s}} $ = 13 TeV}},
  \href{https://doi.org/10.1007/JHEP11(2024)036}{\emph{JHEP} {\bfseries 11}
  (2024) 036} [\href{https://arxiv.org/abs/2407.09183}{{\ttfamily
  2407.09183}}].

\bibitem{corpe_2024_12957031}
L.D.~Corpe and H.~Abdelhamid, ``{Re-interpretation BDTs for
  ATLAS-EXOT-2022-04}.'' DOI:
  \href{https://doi.org/10.5281/zenodo.12957030}{10.5281/zenodo.12957030},
  July, 2024.

\bibitem{haddad:ramp}
A.~Haddad, ``{ATLAS-EXOT-2022-04: Using a BDT to recast Displaced hadronic jets
  with leptons or extra jets}.'' {RAMP seminar, 16 Oct 2024,
  \url{https://indico.cern.ch/event/1463907/}}.

\bibitem{Dreyer:2024bhs}
E.~Dreyer, E.~Gross, D.~Kobylianskii, V.~Mikuni, B.~Nachman and N.~Soybelman,
  \emph{{Automated Approach to Accurate, Precise, and Fast Detector Simulation
  and Reconstruction}},
  \href{https://doi.org/10.1103/PhysRevLett.133.211902}{\emph{Phys. Rev. Lett.}
  {\bfseries 133} (2024) 211902}
  [\href{https://arxiv.org/abs/2406.01620}{{\ttfamily 2406.01620}}].

\bibitem{Heinrich:2021gyp}
L.~Heinrich, M.~Feickert, G.~Stark and K.~Cranmer, \emph{{pyhf: pure-Python
  implementation of HistFactory statistical models}},
  \href{https://doi.org/10.21105/joss.02823}{\emph{J. Open Source Softw.}
  {\bfseries 6} (2021) 2823}.

\bibitem{ATL-PHYS-PUB-2019-029}
{\scshape ATLAS} collaboration, \emph{{Reproducing searches for new physics
  with the ATLAS experiment through publication of full statistical
  likelihoods}},  Tech. Rep.
  \href{https://cds.cern.ch/record/2684863}{ATL-PHYS-PUB-2019-029}, CERN,
  Geneva (2019).

\bibitem{atlas:publicresults}
\url{https://twiki.cern.ch/twiki/bin/view/AtlasPublic}.

\bibitem{Elmer:2023wtr}
N.~Elmer, M.~Madigan, T.~Plehn and N.~Schmal, \emph{{Staying on Top of
  SMEFT-Likelihood Analyses}},
  \href{https://arxiv.org/abs/2312.12502}{{\ttfamily 2312.12502}}.

\bibitem{CMS:2024onh}
{\scshape {CMS}} collaboration, \emph{{The CMS Statistical Analysis and
  Combination Tool: Combine}},
  \href{https://doi.org/10.1007/s41781-024-00121-4}{\emph{Comput. Softw. Big
  Sci.} {\bfseries 8} (2024) 19}
  [\href{https://arxiv.org/abs/2404.06614}{{\ttfamily 2404.06614}}].

\bibitem{cms_collaboration_2024_c2948-e8875}
{CMS Collaboration}, ``{CMS Higgs boson observation statistical model}.'' DOI:
  \href{https://doi.org/10.17181/c2948-e8875}{10.17181/c2948-e8875}, Apr.,
  2024.

\bibitem{Sekmen:2025hwq}
S.~Sekmen, \emph{{Publishing full statistical models of CMS physics analyses}},
  \href{https://doi.org/10.22323/1.476.0988}{\emph{PoS} {\bfseries ICHEP2024}
  (2025) 988}.

\bibitem{hs3:github}
{\scshape {HS3}} collaboration.
  \url{https://github.com/hep-statistics-serialization-standard/hep-statistics-serialization-standard}.

\end{thebibliography}

\providecommand{\href}[2]{#2}\begingroup\raggedright\endgroup

\end{document}